\documentclass[12pt,preprint]{aastex}

\usepackage{graphics,graphicx}
\usepackage{amsmath}
\usepackage{natbib}
\usepackage{color}
\def\vv{{\mathbf{v}}}
\def\vr{{\mathbf{r}}}
\def\vs{{\mathbf{J}}}
\def\vh{{\mathbf{h}}}

\newcommand{\gtorder}{\mathrel{\raise.3ex\hbox{$>$}\mkern-14mu
            \lower0.6ex\hbox{$\sim$}}}
\newcommand{\ltorder}{\mathrel{\raise.3ex\hbox{$<$}\mkern-14mu
            \lower0.6ex\hbox{$\sim$}}}

\shorttitle{Sound Speed Dependence of Tilted Disks}
\shortauthors{Hawley \& Krolik}

\begin{document}

    \title{Sound Speed Dependence of Alignment in Accretion Disks Subjected to Lense-Thirring Torques}

\author{John F. Hawley\altaffilmark{1}}

\and

\author{Julian H. Krolik\altaffilmark{2}}

\altaffiltext{1}{Department of Astronomy, University of Virginia, Charlottesville VA 22904, USA}

\altaffiltext{2}{Department of Physics and Astronomy, Johns Hopkins University, Baltimore, MD 21218, USA} 

\begin{abstract}
We present a series of simulations in both pure hydrodynamics (HD) and magnetohydrodynamics (MHD) exploring the degree to which alignment of disks subjected to external precessional torques (e.g., as in the ``Bardeen-Petterson" effect) is dependent upon the disk sound speed $c_s$.  Across the range of sound speeds examined, we find that the influence of the sound speed can be encapsulated in a simple ``lumped-parameter" model proposed by \citet{SKH13b}.  In this model, alignment fronts propagate outward at a speed $\simeq 0.2 r \Omega_{\rm precess}(r)$, where $\Omega_{\rm precess}$ is the local test-particle precession frequency.  Meanwhile, transonic radial motions transport angular momentum both inward and outward at a rate that may be described roughly in terms of an orientation diffusion model with diffusion coefficient $\simeq 2c_s^2/\Omega$, for local orbital frequency $\Omega$.  The competition between the two leads, in isothermal disks, to a stationary position for the alignment front at a radius $\propto c_s^{-4/5}$.  For alignment to happen at all, the disk must either be turbulent due to the magnetorotational instability in MHD, or, in HD, it must be cool enough for the bending waves driven by disk warp to be nonlinear at their launch point.  Contrary to long-standing predictions, warp propagation in MHD disks is diffusive independent of the parameter $c_s/(\alpha v_{\rm orb})$, for orbital speed $v_{\rm orb}$ and ratio of stress to pressure $\alpha$. In purely HD disks, i.e., those with no internal stresses other than bulk viscosity, warmer disks align weakly or not at all; cooler disks align qualitatively similarly to MHD disks.
\end{abstract}


\section{Introduction}

Considerable theoretical effort has been devoted to understanding disk alignment, but progress has been slow. One reason stems from one of the earliest results in the field.  As shown long ago by \citet{PP83}, warped disks necessarily create radial pressure gradients.  These radial pressure gradients then induce radial fluid motions, which can carry differently-aligned angular momentum from one radius to another.  One of the central questions to answer is therefore the nature of the mechanisms governing the speed of these radial motions.  Reynolds stresses result from these motions, but they are difficult to quantify  because they correspond to different elements of the stress tensor than those associated with accretion.

 Early efforts to define these stresses focused on analytical approaches \citep{BP75,Hatchett81,PP83,Pringle92,PapLin95}.  It was therefore natural to adopt a simple prescription for stresses, the \citet{SS73} ``$\alpha$" model.  However, applying this model to warped disks required extending it from its original definition (vertically-integrated and time-averaged stresses mediating accretion in flat disks, not necessarily tied to any particular mechanism) to one in which it was specifically a viscosity (therefore negatively proportional to shear) and applied to all components of the stress tensor.  Introduction of this {\it ansatz} led to a dichotomy in warped disk problems, dividing them according to whether the ratio of stress to pressure $\alpha$ was greater than the disk aspect ratio (the ``diffusive" regime) or the other way around (the ``bending wave" regime) \citep{PP83,PapLin95}.

 Another reason for slow progress is the centrality of nonlinear fluid dynamics, for which numerical simulation is a better tool than analytic methods.  An exemplar is the pioneering work of \cite{NP00}, which investigated nonlinear effects in the alignment process through numerical simulations using smoothed particle hydrodynamics (SPH).  In their calculation, internal stress was provided by the intrinsic numerical viscosity of the SPH code; this effective viscosity was calibrated with a series of bending wave calculations \citep{NP99}.  They examined alignment for different disk thicknesses (e.g., Mach number $=12$, 30), different black hole inclinations ($10^\circ$, $30^\circ$), and a variety of Newtonian and pseudo-Newtonian potentials designed to model the Lense-Thirring effect; in some cases, they also included relativistic apsidal precession.  They found, among other things, that the alignment transition radius $r_T$, defined as the point where the disk tilt has a value halfway between the initial misalignment and full alignment, lies at smaller radius than originally predicted by \cite{BP75}, and that $r_T$ is larger for a thinner disk.  They proposed several analytic models, all of them based on versions of the diffusion approximation, for the location of $r_T$ as a function of disk and black hole parameters.  Since this work, there have also been numerous other SPH-based investigations, all assuming an $\alpha$ model, {\it viz.} that the fluid's internal stresses can be described by an isotropic viscosity whose stress is linearly proportional to the local pressure \citep{LP07,Lodato10,Nixon2012b,Nealon2016}.

Internal stress is an essential component of disk dynamics, but it does not originate with some unknown $\alpha$ viscosity, nor is it necessarily isotropic; rather, it arises from MHD turbulence within the disk that is driven by the magneto-rotational instability \citep[MRI;][]{mri91,bh98}. MHD simulations require abandoning SPH as the numerical technique and adopting a grid-based scheme.  This is not without its difficulties, however.  To capture the MHD turbulence, a simulation must  have relatively fine resolution within the disk (dozens of grid zones across a pressure scale height $h$) and a timestep very short compared to an orbital timescale.   As a further challenge, the Lense-Thirring precession frequency $\Omega_{precess}$ at the transition radius is considerably smaller than the orbital frequency $\Omega$.   Consequently, simulations able to probe the alignment transition at a realistic scale are costly.   One way to address this challenge is to focus on the near-hole region, where the various time-scales are not too dissimilar, but if \cite{NP00} are correct, the disk must be geometrically thick for the alignment transition to occur close to the black hole.   \cite{Fragile07} and \cite{Fragile14a} investigated this region with MHD simulations of mis-aligned disks in the Kerr metric.  A problem with working so close to the black hole, however, is that the inflow rate is rapid near the innermost stable orbit (the ISCO), both because the ISCO is not far inside this region and because the disk's thickness implies a relatively high pressure, and radial pressure gradients accelerate inflow.  As a result, matter can be carried into the black hole faster than the rate at which the disk can precess or align.  \cite{Liska2018} carried this program a step further, focusing on the interaction between the alignment of thick disks and jets.

We have adopted a different approach.  Rather than considering the problem  with a relativistic metric,  we instead include only a lowest-order post-Newtonian term to represent the Lense-Thirring torque in order to focus on idealized models in which the physical processes can be studied in isolation and in detail.  Thus, we are not trying to simulate realistic disks, nor would we expect the specific results of any one simulation to be found in Nature; instead the goal is to isolate the principal mechanisms at work.   Moreover, throughout this program we consider only genuinely physical mechanisms: that is, we eschew any use of a phenomenological viscosity.

We began with the simulations of \cite{SKH13a}, who considered the simple case of the relaxation of an unforced warped disk in pure hydrodynamics, without internal stresses.  Angular momentum transport was controlled by unbalanced pressure gradients associated with the warp. \cite{SKH13a} argued on the basis of their simulations that the most relevant distinction is actually between linear and nonlinear bending waves.  If the local direction of the angular momentum is defined by the unit vector $\hat\ell$, bending waves become nonlinear when the induced radial pressure contrast across an $e$-fold in radius is order unity, i.e., when $\hat\psi \equiv | d\hat\ell/d\ln r|/(h/r) > 1$.  Linear waves can propagate through a laminar background;  nonlinear waves are damped quickly as they induce shocks \citep{NP99,SKH13a}.

Our next study \citep{SKH13b} simulated a misaligned disk with fully-developed MHD turbulence in Newtonian dynamics. To represent the Lense-Thirring (gravito-magnetic) torque, we follow \citet{NP00} and use the form $\rho \vv \times \vh$, where $\rho$ is the mass density, $\vv$ is the fluid velocity, and
\begin{equation}\label{eq:pnterm}
\vh = \frac{2\vs}{r^3} - \frac{6(\vs \cdot \vr)\vr}{r^5}.
\end{equation}
Here $\vs$ represents the specific angular momentum of the central mass and $r$ is spherical radius; the parameter $a$ used to describe spin in a Kerr spacetime specifies its magnitude in units of $r_g c$, where $r_g = GM/c^2$.  In terms of this parameter, the Lense-Thirring precession frequency is
\begin{equation}
\Omega_{precess} = \frac{2G^2Ma}{r^3c^2}.
\end{equation}
 Because $\Omega_{precess} \ll \Omega$ except very near the black hole, and numerical simulation of MHD turbulence demands resolving timescales $\ll \Omega^{-1}$, the computational expense of such a simulation can be prohibitive.  To alleviate this problem while retaining the essential physics, we chose to multiply $\Omega_{precess}$ by a constant factor large enough to make it $\sim O(0.1) \Omega$: sufficiently small compared to the dynamical frequency for orbital dynamics to dominate, but not so small as to make the simulations impossible.

Making this parameter adjustment, \cite{SKH13b} evolved a moderately thick ($h/r \sim 0.1$--0.2) disk with an adiabatic equation of state, using both hydrodynamics (HD) and MHD.  The results showed no support for the assumption that an isotropic viscosity limits vertical shear, as the actual magnitude of the $r$--$z$ component of the magnetic stress is both very small and carries a sign uncorrelated to the sign of the shear (unlike any sort of viscous stress or the $r$--$\phi$ stress resulting from the MRI).  Thus, what regulates the magnitude of the radial velocities is {\it not} viscosity, but pressure gradients and gravity. Because so much previous work had assumed the existence of an ``isotropic $\alpha$ viscosity", this finding calls into question much that had seemed well-established.  In particular, if this viscosity does not actually exist, what is the meaning of a regime distinction based upon it?

This work also emphasized a prerequisite for disk alignment: a negative precession phase gradient,  so that the angular momentum carried outward by the radial motions has the correct sign for alignment.  It further found that although magnetic forces are in general weaker than pressure forces in warped disks, MHD turbulence disrupts the phase coherence of bending waves, delaying the enforcement of solid body precession and maintaining the precession phase gradient.  MHD turbulence also completes alignment even when the remaining warp is too small to drive Reynolds stresses. This MHD simulation did not, however, achieve a steady state transition front; the alignment front traveled outward to where the available gas, and hence misaligned angular momentum, began to rapidly decline.   Strikingly, the paired HD simulation failed to align, an indication of the importance of MHD turbulence to the alignment process, even though its direct contribution to alignment is at most secondary.

\cite{KH2015} examined a disk with an isothermal equation of state in which the surface density increased with radius. The alignment front moved outward through the disk, but slowed, reversed and then stopped, illustrating how a steady-state transition front could be formed.   To explain the alignment front propagation speed, $dr_f/dt$, and the radius of the steady state transition radius $r_T$, \cite{SKH13b} and \cite{KH2015}  proposed that the propagation speed of the alignment front is determined by the rate at which angular momentum whose direction could cancel the misalignment could be carried outward in the disk.  This rate is characterized by the angle $\gamma$ between the angular momentum (perpendicular to the black hole spin axis) being carried outward and the direction opposite to the local misaligned angular momentum (here we use ``local" to mean ``averaged on a spherical shell").  The transported angular momentum optimally cancels the misaligned angular momentum when $\gamma = 0$.  The local torque scales with the surface density $\Sigma$ and $\sin\beta$, the local misalignment.  The alignment front propagation speed is the ratio of this torque to the local misaligned angular momentum, and is given by
\begin{equation}\label{eq:rfspeed2}
{d r_f\over dt} = \langle \cos\gamma\rangle \mathcal {I}(r_f) r_f \Omega_{precess},
\end{equation}
where the averaging over $\cos\gamma$ refers to an average over the turbulence.
${\mathcal I}$ is the dimensionless integral
\begin{equation}\label{eq:scrI}
\mathcal{I}(r)   = \int_0^1 \, dx \, x^{-3/2} \frac{\sin\beta(x)}{\sin\beta(r_f)}
   \frac{\Sigma(x)}{\Sigma(r_f)},
\end{equation}
in which $x = r/r_f$.

 \cite{SKH13b} and \cite{KH2015} proposed that alignment stalls where the speed of the alignment front matches the speed with which misaligned angular momentum from the outer disk is mixed inward.  If this inward mixing is described as a diffusion-like process, the diffusion coefficient is $\sim c_s^2/\Omega$.  With this assumption, \cite{KH2015} developed a relation for the transition radius $r_T$ as a function of $h/r$.  Because \cite{KH2015} considered only a single value of $h/r$, however, they could not fully test this hypothesis.

Dimensional analysis leads to an effective diffusion coefficient proportional to $c_s^2/\Omega$; consequently, this scaling is shared with earlier diffusion models based on different mechanisms \citep{Pringle92,SF96,NP00}.  However, contrasting underlying mechanisms lead to different predictions for the dimensionless factor multiplying $c_s^2/\Omega$.  Initially, it was thought to be $\sim \alpha^{-1}$ \citep{Pringle92,SF96}.  \citet{O99} then pointed out that this estimate applied only to small amplitude warps and developed a new prediction better-suited to the quasi-linear regime.   As a nonlinear theory, however, it leads to an effective warp diffusion coefficient that varies with position and time.  The SPH simulations \cite{Lodato10}, which used an isotropic alpha viscosity, found consistency with the \cite{O99} predictions for an averaged value of the warp amplitude when $\alpha > 0.2$.  However, for $\alpha \le 0.2$ (the physical regime if $\alpha$ is estimated in a way informed by time-averaged MHD simulations), the actual warp profile was not well-fit, leading to progressively larger uncertainty in the numerical inference of the warp diffusion coefficient as alpha decreased.
Further, even modest warps in this theory lead to negative values of the effective accretion stress. More fundamental questions were raised about the applicability of a diffusion model by \citet{SKH13a}, who found that the stress associated with nonlinear warps in pure non-viscous hydrodynamics was neither linearly proportional to the warp amplitude nor simultaneous with the warp.  In addition, no diffusion model can be successful without a mechanism to maintain the appropriate precession phase gradient interior to the transition front.

The purpose of the present study is to deepen our investigation of  the transition front, and to test further the  applicability of the diffusive model as it appears in our formulation.  The prime question is:  What is the dependence of the alignment process and the location of $r_T$ on the disk sound speed? Unless $r_T$ scales in the right way with $c_s$, {\it no} diffusive model can be correct.  Equating the alignment front speed with the diffusion velocity yields
\begin{equation}
\langle -\cos\gamma\rangle {\mathcal I} r_T \Omega_{precess} = A \left[c_s^2 /(r_T \Omega)\right] B(r_T),
\end{equation}
where $\Omega$ is the local orbital frequency  and $A$ is the dimensionless factor in the diffusion coefficient.  The quantity $B \equiv  |\partial \sin\beta/\partial \ln r| / \sin\beta$ so as to give the characteristic rate at which misaligned angular momentum is transported radially by diffusion.
Inserting the radial dependences for $\Omega_{precess}$ and the orbital frequency, we find
\begin{equation}\label{eq:rT}
r_{T}/r_g = \left[\frac{ 2(a/M) \langle \cos\gamma\rangle {\mathcal I}}{AB(r_T)}\right]^{2/5} (c/c_s)^{4/5}.
\end{equation}
In other words, diffusive models in general predict $r_T \propto c_s^{-4/5}$.  Here we present new isothermal disk simulations, similar to the one carried out in \cite{KH2015}, but with reduced values of $c_s$.   Our aim is to test this scaling across as wide a range of disk aspect ratio as possible (thereby also challenging the $\alpha$-based regime dichotomy) and, if the scaling is confirmed, to calibrate the dimensionless factor $A$.   We will, in addition, study the degree to which diffusion may or may not describe the time-dependent processes leading to an alignment steady-state.

\section{Simulation Methods and Parameters}

\subsection{Model system}.

In the present work, we continue to use the model first studied in \cite{KH2015}, an isothermal disk orbiting a point-mass in Newtonian gravity with a Keplerian angular velocity distribution, $\Omega^2 = GM/r^3$.  Since we employ a Newtonian potential, the radial units are arbitrary, in contrast to both relativistic gravity or a pseudo-Newtonian potential defined in terms of a gravitational radius $r_g = GM/c^2$.   To avoid creating a naked singularity, $|a/M| \leq 1$ for real black holes; by contrast, in our scale-free Newtonian approximation, we can regard $\vs$ in Eq.~(\ref{eq:pnterm}) as a free parameter whose magnitude is unbounded.  The only physical constraint is to preserve the ordering $\Omega_{precess} < \Omega$.  For the simulations in this paper, we set $GM = 1$, and $\Omega_{precess} = 2/r^3$,  which is equal to 1/15.8 of the orbital frequency at the fiducial radius of $r=10$.  We report time in units of fiducial orbits, defined as 200 units of code time, which is almost exactly the orbital period at $r=10$, e.g.,  $P_{\rm orb} = 2\pi r^{3/2} = 199$ at $r=10$.

As in our earlier papers, we omit general relativistic apsidal precession, which causes the epicyclic frequency to differ from the circular orbital frequency.  Because the focus of these studies is on alignment far from the black hole, where the apsidal precession frequency is much slower than the eddy turnover rate in the MHD turbulence, we expect that any effects due to apsidal precession would be minimal.  Relativistic apsidal precession has been modeled in Newtonian HD simulations of warped disks \citep{NP00,Nealon2016}.   Both studies, focusing on distances only a few tens of $r_g$ from the black hole, found that apsidal precession could influence alignment, but their other results contrasted strongly, and neither identified any specific physical mechanism by which apsidal precession modifies alignment.  This question can more fruitfully be explored once the simpler problem is better explicated.

The sound-speed $c_s^2$ of the isothermal equation of state is selected to set the scale height of the disk $h=c_s/\Omega$.  The scale height varies $\propto r^{3/2}$, making the disk aspect ratio flare outward $\propto r^{1/2}$.  We set the density at the equator $\rho_c = 1$ at all radii and determine its vertical distribution by assuming it is in hydrostatic equilibrium, i.e., $\rho = \rho_c \exp (-z^2/2h^2)$.   At the initial inner ($r=6$) and outer disk limits, the disk is truncated.  Consequently, it is not in true radial pressure equilibrium (especially at the disk boundaries), and in the subsequent evolution the disk's outer boundary moves outward from where the disk was initially truncated.
The surface density $\Sigma \propto h$ increases outward $\propto r^{3/2}$ until the outer portion of the disk where, due to the finite size of the disk, $\Sigma$ smoothly declines to zero.  This surface density profile does not correspond to inflow equilibrium for this temperature distribution, but does constitute a simple model where the sound speed is the dominant factor in distinguishing one case from another.  It should be noted that not all disk systems in Nature are necessarily in inflow equilibrium, and in any case alignment generically proceeds faster than inflow \citep{PP83}. 

The initial magnetic field is defined by a vector potential proportional to the square root of the disk density within an ``envelope'' function,
\begin{equation}
A_{\phi} = A_0 \rho^{1/2} \sin \left[{\pi\over 2} ({r_o/ r})^{1/2}\right] (r/r_{in} -1) (1-r/r_{out}) 
\end{equation}
where $r_o =4$ is the grid inner boundary, $r_{in}$ is the disk inner radius and $r_{out}$ is the disk outer radius.
The vector potential is limited to positive values with a cutoff at $0.05\rho_c$, i.e.,
\begin{equation}
A_{\phi} = \max(A_{\phi} - 0.05 \rho_{c},0).
\end{equation}  
The field amplitude factor $A_0$ is chosen so that the initial volume-integrated ratio of gas to magnetic pressure, the plasma $\beta$, is 1000.  This particular vector potential leads to weak, primarily-radial,  magnetic field that rapidly generates toroidal field through Keplerian shear.  Although a radial field is MRI unstable {\citep[][provides an example of the evolution of the radial MRI]{1992ApJ...400..595H}}, the toroidal field MRI proves to be the most significant in generating MHD turbulence.  We seed the MRI by imposing 1\% random pressure perturbations on the initial condition.  

To ensure that the MHD turbulence is well-developed before we study the effects of Lense-Thirring torques, we let the disk evolve from this initial condition until MHD turbulence is developed in the inner disk, typically 15--20 fiducial orbits.  At that point, the torque is turned on and the disk evolved toward an alignment steady-state.
\subsection{Numerics}

As in our previous MHD simulations of warped disks \citep{SKH13b, KH2015}, we use our Fortran-95 version of the 3D finite-difference code {\it Zeus} \citep{zeus1,zeus2,hawleystone95}.  The {\it Zeus} code solves the standard equations of Newtonian MHD (supplemented by the torque term previously described) using direct finite differencing.  We work in spherical coordinates $(r,\theta,\phi)$.  The radial grid extends outward from a minimum value using a logarithmically graded mesh.  Because we are working with relatively thin disks, and we wish to avoid potential difficulties with coordinate singularities near the axis, we limit the extent of $\theta$ to the interval $[0.1,0.9]\pi$.  The $\theta$ zones are concentrated around the equatorial plane using the polynomial spacing given by equation (6) of  \cite{NKH10}, 
\begin{equation}\label{eqn:grid}
\theta(y) = {\pi \over 2}\left[ 1+(1-\xi)(2y-1) + (\xi - {2\theta_c \over \pi})(2y-1)^n \right]
\end{equation}
The $\theta$ grid index is $y= (i+0.5)/N$, where $i$ is the zone-index and $N$ is the total number of $\theta$ zones; $\theta_c$ is the size of the ``cutout'' around the polar grid axis, $\xi = 0.65$, and $n = 13$.  The resulting distribution of zones has a relatively large $\Delta \theta$ near the cutouts along the axis, smoothly decreasing to a small, constant $\Delta \theta$ over a symmetrical region surrounding the equator. The $\phi$ coordinate covers the full $2\pi$ in angle with uniform spacing.

\subsection{Diagnostics}

As in our previous studies, we establish Cartesian coordinates to describe how the disk tilts and warps, choosing the direction of the black hole spin $\vs$ to define the $z$ axis.  The polar axis of the code's spherical grid, which is parallel to the initial disk angular momentum, is in the $x$-$z$ plane, tilted $12^\circ$ (0.21 radians) from the $z$-axis in the $\hat x$ direction.  For questions of alignment and evolution of the disk angular momentum, we map the disk's angular momentum vector onto this Cartesian coordinate system.  At each radius we compute a shell average of the angular momentum, $\vec \ell (r)$, and transform the resulting averaged vector into the Cartesian system.  We can then define several diagnostic quantities, such as the misalignment angle
\begin{equation}\label{eq:beta}
\beta = \tan^{-1}\left(|\ell_\perp|/|\ell_z|\right),
\end{equation}
where $\ell_\perp^2 = \ell_x^2+\ell_y^2$.  In all the simulations reported here, the value of $\beta$ starts at $12^\circ$; at perfect alignment $\beta = 0$.  The precession angle is defined as
\begin{equation}\label{eq:phi}
\phi_{prec} = \tan^{-1}\left(\ell_y/\ell_x\right);
\end{equation}
$\phi$ increases as the disk angular momentum vector precesses around the $z$-axis.

To gauge the numerical quality of the simulation we employ certain metrics previously developed and studied in \cite{HGK11,HRGK13}, as well as in \cite{Sorathia12}.  The $Q$ metrics measure the number of grid cells spanning a characteristic MRI wavelength $\lambda_{MRI}= 2\pi |v_{A}|/\Omega$.  Their specific definitions are
\begin{equation}\label{eq:qtheta}
Q_\theta = {\lambda_{MRI\theta} \over r \Delta \theta},
\end{equation}
and
\begin{equation}\label{eq:qphi}
 Q_\phi = {\lambda_{MRI\phi}\over r \sin\theta \, \Delta\phi }.
\end{equation}
The  Alfv\'en speed $v_A$ in the expression for the characteristic MRI wavelength is obtained from the appropriate component of the magnetic field ($B_\theta$ for $Q_\theta$, $B_\phi$ for $Q_\phi$).  The $Q$ metrics are density-weighted shell averages; larger values tend to yield more fully-developed turbulence.  \citet{HGK11} and \citet{HRGK13} estimate that $Q$ values $>15$--$20$ are indicative of adequate resolution.  

A second set of metrics measures the average properties of fully developed MHD turbulence; these metrics are calibrated by the values measured in highly-resolved local simulations.  \cite{HGK11,HRGK13} developed two such diagnostics: $\alpha_{mag} = M_{r\phi}/P_{mag}$, the ratio of the Maxwell stress, $M_{r\phi}=-B_r B_\phi/4\pi$, to the magnetic pressure; and $\langle B_r^2\rangle/\langle B_\phi^2 \rangle$, the ratio of the radial to toroidal magnetic energy.  When suitably averaged over the computational domain in well-resolved simulations, including local shearing box simulations, these quantities are $0.45$ and $0.2$, respectively \citep{HRGK13}. 

We can also characterize the magnetic stress in the disk in terms of the traditional $\alpha$ parameter.  Here we define a shell-averaged $\alpha$ as
\begin{equation}\label{eq:alpha}
{\alpha}\left (r \right) = {1 \over 2\pi} \int \, d\phi \, {\int M_{r\phi} r d\theta \over \int \rho c_s^2 r d\theta }.
\end{equation}
It should be noted that both stress and density vary locally and along all axes, and that, for simplicity, we compute the stress in terms of the original $(r,\theta,\phi)$ grid, even when the disk is aligned with the black hole equatorial plane.  The small tilt angle we assume ($12^\circ$) makes that a reasonable approximation.

\subsection{Simulations}

\begin{deluxetable}{lccccc}
\tabletypesize{\scriptsize}
\tablewidth{0pc}
\tablecaption{Simulation List\label{table:list}}
\tablehead{
  \colhead{Name}&
  \colhead{$(r,\theta,\phi)$}&
  \colhead{$c_s^2$}&
  \colhead{$h/r(=10)$} &
  \colhead{Orbits}&
  \colhead{$r_{\rm out}$} 
  }
\startdata
KH-2015      & $352\times 384\times 1024$  &$10^{-3}$   & 0.1 & 12.4 & 35  \\
KH-2015-H  & $352\times 384\times 1024$  &$10^{-3}$  & 0.1 & 11.1 & 35 \\
High-thin     & $704\times 770\times 1024$  &$2.5\times 10^{-4}$& 0.05 &   22.3 &  31 \\
Low-thin      & $320\times 400\times 500$    &$2.5\times 10^{-4}$& 0.05 & 22.6 &  28  \\
Thin-H  & $320\times 400\times 500$    &$2.5\times 10^{-4}$& 0.05 &  33.8 &  28   \\
V-thin          & $765\times 765\times 1024$  &$1.25\times 10^{-4}$&0.035 & 15.2 & 37  \\
V-thin-H      & $400\times 400\times 400$    &$1.25\times 10^{-4}$&0.035 & 21.0&  35\\
Big-H          & $600\times 400\times 400$    &$2.5\times 10^{-4}$&0.05 &  20.8 & 100    
\enddata
\end{deluxetable}

Table \ref{table:list} lists the simulations carried out for this study along with the simulation from \cite{KH2015}.  The table gives: the name of the simulation; the number of grid-cells employed; the sound speed;  $h/r$ at the fiducial radius $r=10$; the run duration with torque in units of fiducial orbits; and the radius of the outer boundary of the disk at the onset of the LT torque.  This last quantity is defined as the radius where the azimuthally-averaged surface density drops below 5\% of the initial maximum value, except for Big-H, where it is the outer grid boundary.  For each MHD simulation, we also ran a purely hydrodynamic counterpart.  These models are labeled with a suffix ``H''.

Our previous simulation, KH2015, was an isothermal disk with $c^2=0.001$, so that $h=c_s/\Omega = 1$ at the fiducial radius $r=10$ (i.e., $h/r = 0.1$ at that location). This simulation used a grid of $(352, 384, 1024)$ zones in $(r,\theta,\phi)$, spanning a range of $[4, 40]$ in radius, $[0.1, 0.9]\pi$ in $\theta$ and the full $2\pi$ in $\phi$.  As the primary goal of the present work is to examine the influence of sound speed in the alignment of the disk, we will contrast KH2015 with two thinner disks.

The High-thin model has half the sound speed and therefore half the scale height of KH2015, with $h/r = 0.05$ at the fiducial radius.  The reduction in scale height demands a greater number of grid cells if the number of zones per $h$ is to be maintained.  It uses $(704, 770, 1024)$ cells in $(r,\theta,\phi)$.  For comparison purposes we also carried out a lower resolution version of this disk (Low-thin), which has $(320, 400, 500)$ cells.  The radial mesh is spaced logarithmically between $r=4$ and 35 in the low resolution simulation and between $r= 4$ and 40 in the high resolution case.  In the high resolution simulation, $\Delta r = 0.0327$ at the fiducial radius, and the $\theta$ zones are concentrated around the equatorial plane, as described above;  $\Delta \theta = 0.0014$ in the plane, corresponding to about 36 zones per $h$ ($= 0.5$ at the fiducial radius).  In the lower resolution simulation, $\Delta r$ at $r=10$ is 0.068, and the minimum $\Delta \theta$ is 0.0027.  The $\phi$ zones are always uniform in size, $\Delta \phi = 0.0061$ in High-thin, 0.0126 in Low-thin.  Outflow boundary conditions are employed on the radial inner and outer boundaries, and along the $\theta$ boundary that forms a ``cut-out'' around the polar axis.

The initial evolution phase of High-thin was computed without any applied LT torque and lasted 18~orbits.  At the end of the initial ``no torque'' phase (18 orbits), $Q_\phi$ rises from 10 at $r=5$ to values above 30 for $r > 9$.  $Q_\theta$ has a similar profile, but is only $\approx 0.85 Q_\phi$.  These values indicate that the primary MRI wavelength is well-resolved. Low-thin is similar, but with $Q$ values about half those in High-thin.
At the end of the ``no torque" evolution in High-thin, $\alpha_{mag}$ had an average value of 0.28 and $\langle B_r^2\rangle/\langle B_\phi^2 \rangle$ had an average value of 0.14 between $r=5$ and $25$ (the main portion of the disk).   These are below the values associated with well-developed MHD turbulence in highly resolved shearing sheet simulations \citep{HRGK13}, but they are relatively constant across the radial range of the disk.      Following this initial evolution the LT torque was turned on, and the simulation ran for an additional 22.3 orbits.  

In our second comparison run, V-thin, the sound speed is reduced by a further $\sqrt 2$, with $c_s^2 = 1.25\times 10^{-4}$, making $h/r = 0.035$ at $r=10$.  Resolution requirements make it challenging to go to much smaller sound speeds, at least with a single-grid system. V-thin uses $765\times 765\times 1024$ grid zones; the radial mesh is logarithmically distributed between $r=4$ and 45, with $\Delta r = 0.032$ at the fiducial radius.  The $\theta$ zones are distributed according to Eq. (\ref{eqn:grid}) with a minimum $\Delta \theta = 0.0014$, corresponding to 25 zones per $h$ at the fiducial radius.  The $\phi$ zones are spaced identically as those in High-thin.
As with the other models, this disk was initially evolved without any applied torque, in order to allow MHD turbulence to develop.  To save computational time, however, the evolution without torque was limited to a $\phi$ domain $\left[ 0, \pi/2\right]$.  Because the primary MRI wavelengths are considerably smaller than $r\pi/2$, and the resulting turbulence is local, this should be adequate to establish the initial disk.  The no-torque disk was evolved for 17.25 orbits, after which the data were replicated over the remaining azimuthal quadrants, small random perturbations were added to break $m=4$ symmetry, the torque was switched on, and the simulation was run for an additional 15.2 orbits. At the end of the initial ``no torque'' phase,  $Q_\theta$ varies between 10 and 25 inside of $r=11$, rising to above 30 for $r > 12$.  $Q_\phi \sim 20$ inside $r=20$, declining slowly with radius beyond that point.  These values are somewhat at the low end of adequate resolution.
At the end of the ``no torque" evolution, $\alpha_{mag}$ had an average value of 0.18 and $\langle B_r^2\rangle/\langle B_\phi^2 \rangle$ had an average value of 0.12 between $r=5$ and $20$.   Again, these are below the values associated with well-developed MHD turbulence in highly resolved shearing sheet simulations \citep{HRGK13}, and lower than the values found in High-thin.

Although alignment depends on the magnitude and properties of the internal stress, much of the process is purely hydrodynamic, i.e., it is principally determined by pressure gradients and gravity \citep{SKH13b}.   Simulation of hydro disks therefore provides a mechanism to investigate certain facets of the alignment process in isolation, as well as to help identify the specific role of MHD stresses \citep{SKH13b}. These purely hydrodynamic (HD) disks are inviscid; there is no internal stress (MHD or viscous), that is, $\alpha = 0$.  Such disks are also of interest in their own right as they can represent weakly magnetized or highly resistive disks, e.g., cold protostellar disks.

Because there is no need to resolve the MRI or the resulting turbulence, HD models require less resolution and are less expensive computationally.  For simulation KH2015-H, however, we chose to use the same initial disk and grid as the original MHD model and simply turned off the magnetic terms at the time when the LT torque was applied.  Any residual internal disk turbulence died out promptly.  The same procedure was used for Thin-H, which began from the Low-thin initial state.  For hydrodynamic models with a different grid from the MHD model, the initial disk was evolved without torque in axisymmetry until the acoustic transients died down.  Subsequently, the disk was mapped to full 3D and then subjected to LT torque.  

\section{Results}

\begin{figure}[h]
\begin{center}
\includegraphics[width=0.5\textwidth]{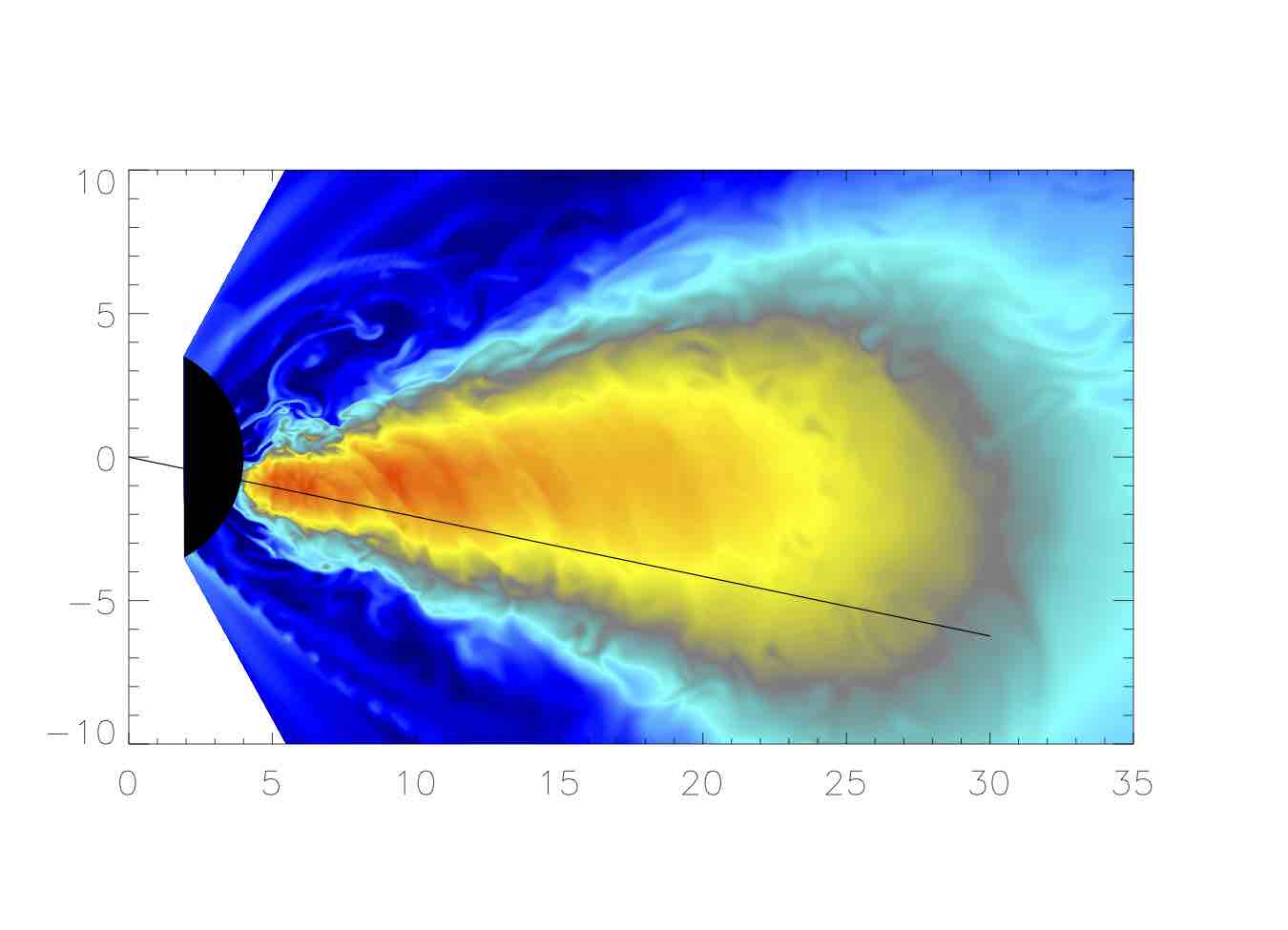}
\vskip -0.3 truein
\includegraphics[width=0.5\textwidth]{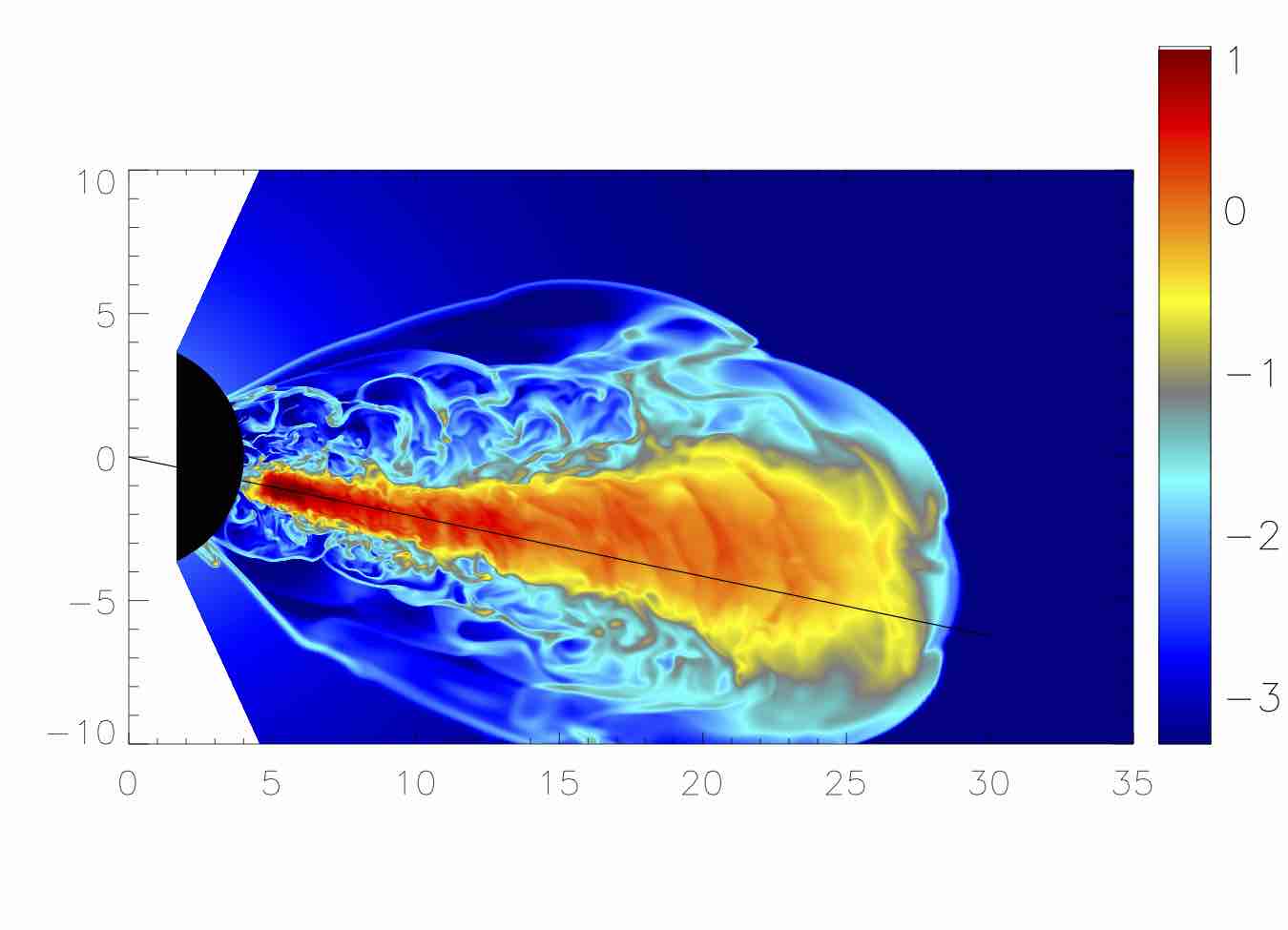}
\vskip -0.3 truein
\includegraphics[width=0.5\textwidth]{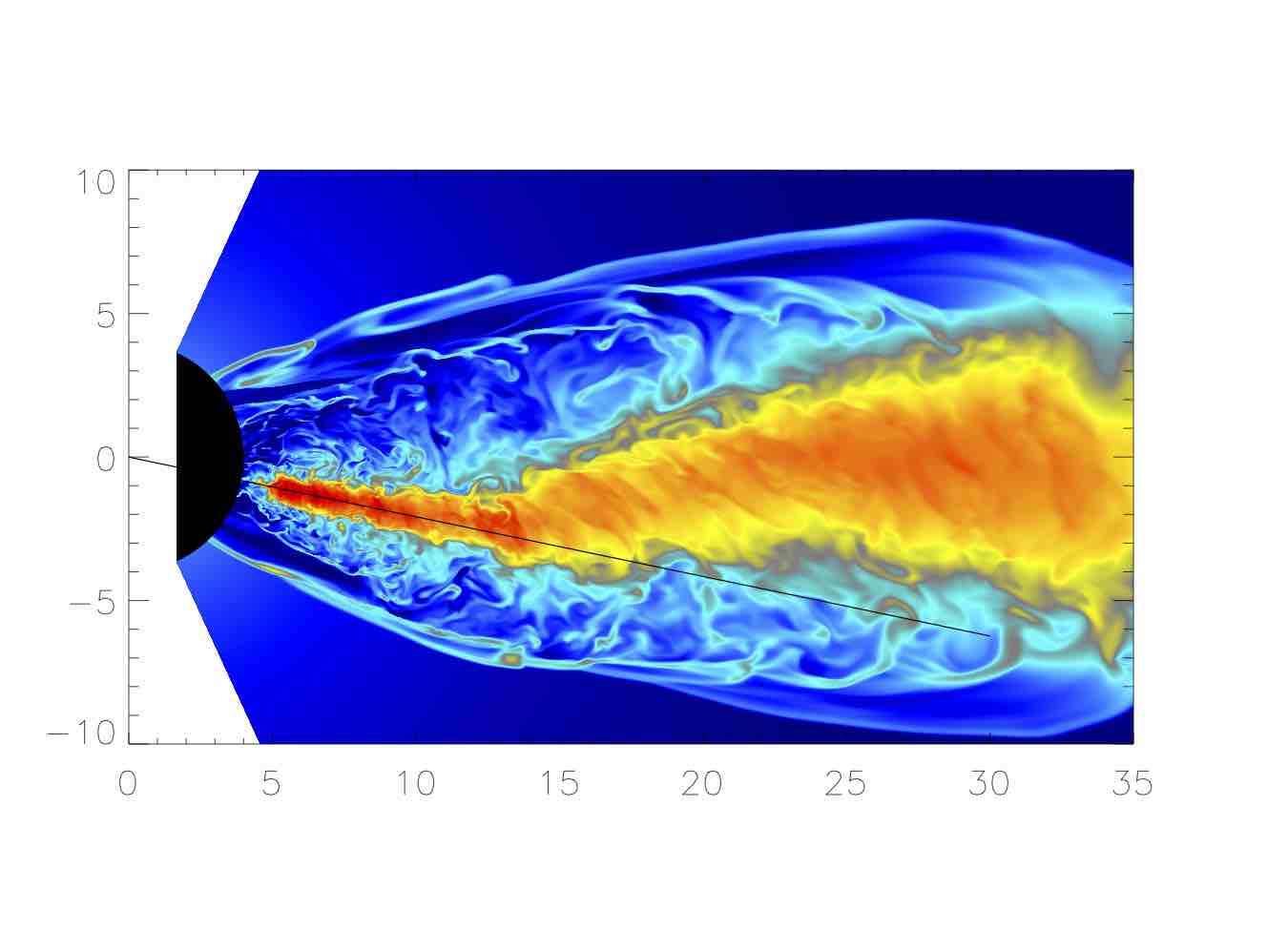}
\caption{Contour plots of of the log density in the $\phi=0$ plane at orbit 12 in KH2015 (top), orbit 24.3 in model High-thin (middle) and orbit 15.2 in model V-thin (bottom). The range in log density is  from 1.0 to -3.3.   Overlaid on each figure is a line showing the equatorial plane for the black hole spin axis of $12^\circ$. }
\label{fig:highdensity}
\end{center}
\end{figure}

Figure~\ref{fig:highdensity}, which shows a density slice for each of the three high-resolution MHD simulations  at late times in their torqued evolution,  gives an introduction to our results.  The density slice is in the $\phi=0$ plane, which corresponds to the maximum tilt of the black hole equatorial plane with respect to the initial disk equator (and grid equator).  The relative thickness of the disks is immediately apparent, as is the greater extent of the aligned region as the sound speed decreases.   In the following subsections, we will elaborate more quantitatively on how the pace and degree of alignment do (or do not) depend on sound speed.

\subsection{Dependence of alignment on sound speed}

\begin{figure}[h]
\begin{center}
\includegraphics[width=0.5\textwidth]{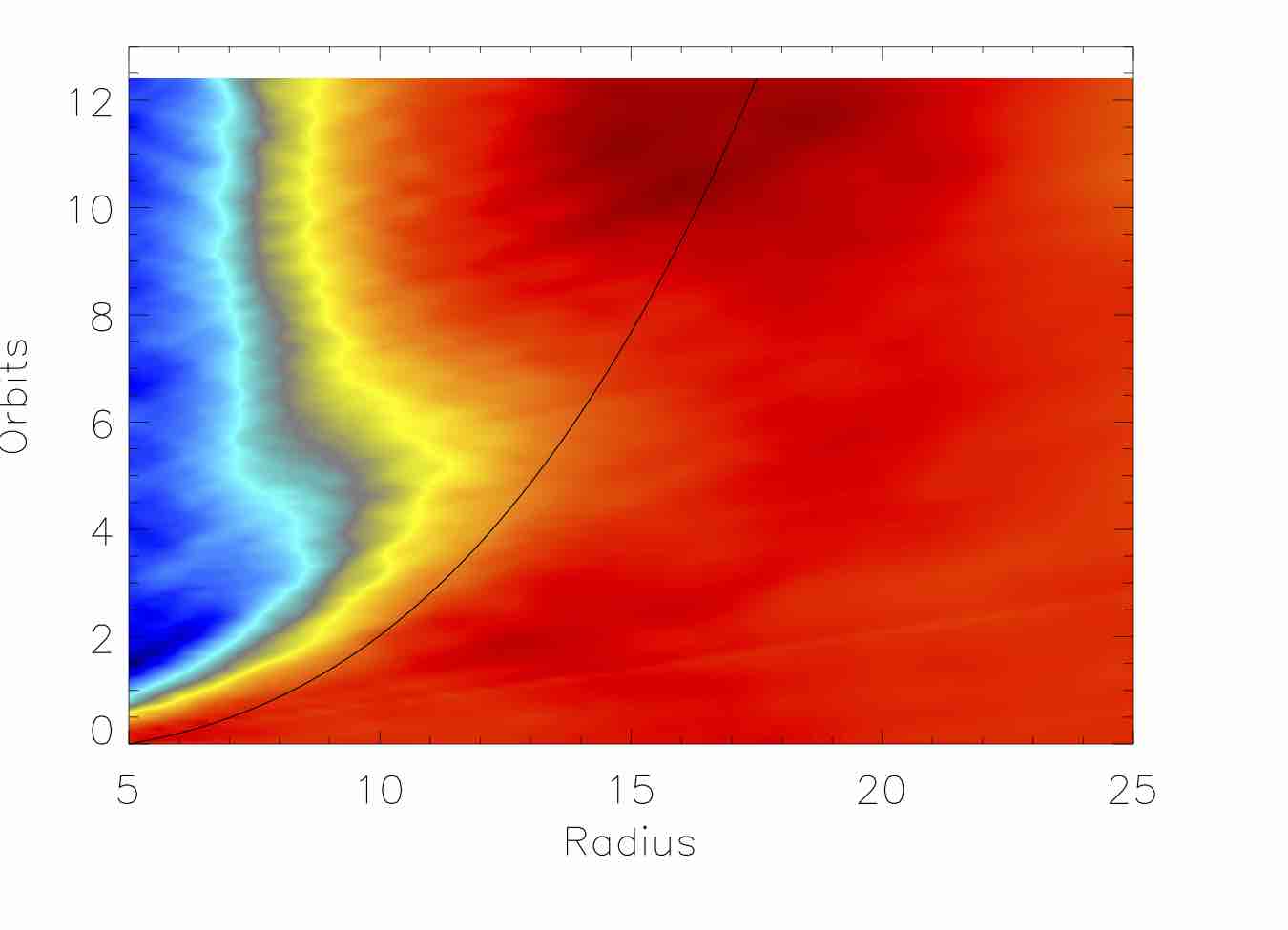}
%\vskip -0.3 truein
\includegraphics[width=0.5\textwidth]{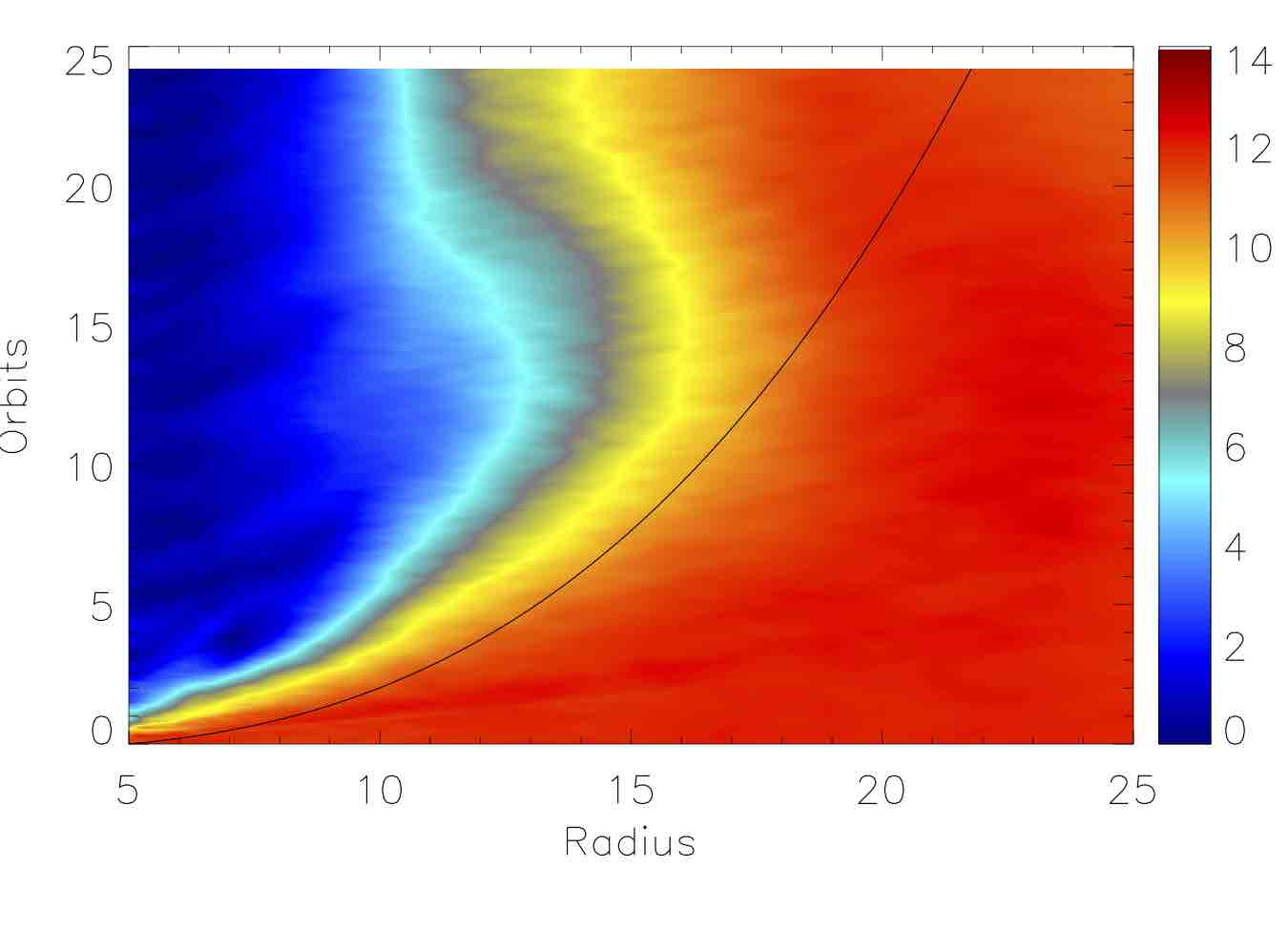}
%\vskip -0.3 truein
\includegraphics[width=0.5\textwidth]{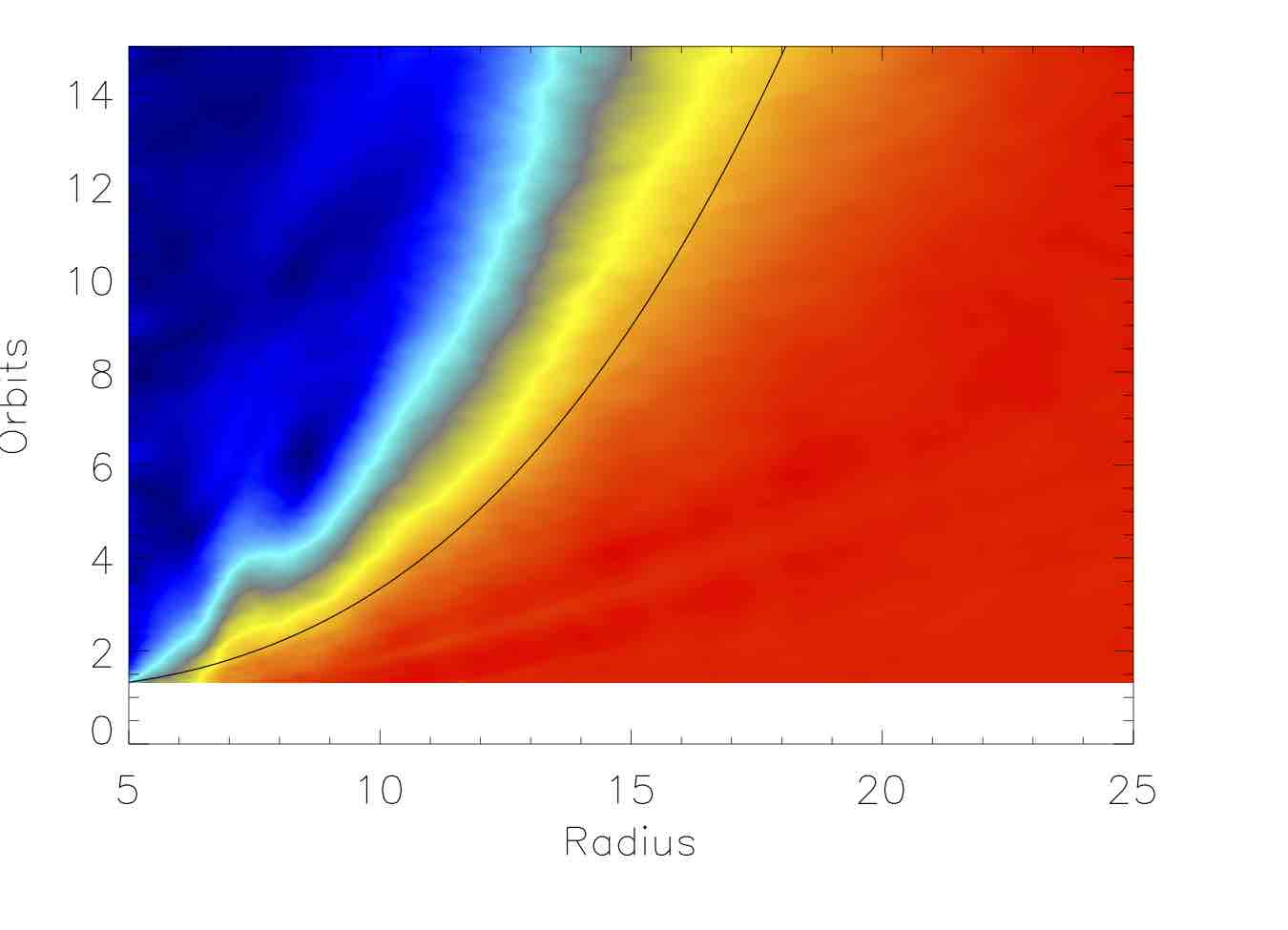}
\caption{Spacetime diagrams (quantities integrated over spherical shells as functions of radius and time) for the alignment angle $\beta$ in simulations KH2015 (top), High-thin (middle), and V-thin (bottom).  Time is in orbits at $r=10$ after the torque is applied. Colors run from $\beta=0$ (blue: aligned) to $\beta=14^\circ$ (red).  Overlaid on the spacetime diagrams is a curve corresponding to an alignment front speed of $ 0.35 r\Omega_{precess}$.  Note that the three simulations ran for different durations (High-thin ran for the largest number of orbits), and the data are missing from the first orbit in the V-thin model. }
\label{fig:highbeta}
\end{center}
\end{figure}

The behavior of the alignment front in the MHD models (KH2015, High-thin, and V-thin) is qualitatively, and in some respects quantitatively, similar (see Fig.~\ref{fig:highbeta}).  In all cases, when the torque is turned on, an alignment front moves out through the disk, and this front initially travels with almost the same speed in all three of these simulations. 
Overlaid on each of the $\beta$ plots in Fig.~\ref{fig:highbeta} is a curve showing a trajectory through spacetime whose radial velocity is $dr_f/dt = 0.35 r \Omega_{precess}(r)$.  We chose the coefficient $0.35$  empirically because it provides a good fit to the progress of the head of the alignment front, defined here as the point where $\beta=10^\circ$ (the red-orange edge in terms of the colorscale).  The alignment front spreads as it moves outward because the front speed decreases where $\beta$ is smaller.  For example, $dr_f/dt = 0.2 r \Omega_{precess}(r)$ where $\beta = 6^\circ$, the half-alignment level.  Within the alignment front the value of $\mathcal I$ (eq.~\ref{eq:scrI}) is $0.3$--0.4, with the lower value during front advancement and $\sim 0.4$ at the stalling radius.  Similar behavior is seen in all three models, except that the alignment front in V-thin continues to advance over the length of the simulation.  During the time when the alignment front moves monotonically outward, these speeds describe the propagation of the alignment front in all three models, regardless of sound speed, suggesting that the proportionality factor does not strongly depend on $c_s$, and that the speed of the alignment front is determined solely by the delivery of aligning momentum by the LT torque for as long as the front is still well inside the stalling radius.  That this should be so is reasonable because the alignment front's  characteristic speed $\propto r\Omega_{precess}$ is both independent of sound speed and  $\propto r^{-2}$, whereas dependence on sound speed enters through the radial mixing speed $\sim c_s^2/(r\Omega)$), which is $\propto r^{1/2}$.  Consequently, the front's motion depends only on the Lense-Thirring angular momentum delivery rate at small radii.

The stalling radius is where those two velocities become comparable, but the front does not stop the first time it reaches this point.  Typically (as seen in the top and middle panels of Fig.~\ref{fig:highbeta}), it overshoots, retreats, and only then approaches its steady-state position.  In both of these cases, where the $\beta =10^\circ$ contour peels away from the model curve is exactly where the alignment front begins to fall back from its maximum overshoot.  Unlike the untrammeled alignment front speed, the stalling radius does depend on sound speed;  we also expect that it will depend on the slope of the alignment profile (eqn.~\ref{eq:rT}).     In KH2015 the front stalls after 5 orbits at about $r=10$, and subsequently retreats to establish a steady state at $r=7$ lasting from orbit 7 until the end of the simulation.  In High-thin the alignment front moves out farther, almost to $r \simeq 15$, before stalling around orbit 12.   After staying at that radius for several orbits, the alignment front subsequently retreats to $r = 12$, where it remains for the last few orbits of the simulation.  In V-thin the alignment front moves outward very nearly monotonically up until the end of the simulation at 15 orbits, when it has reached $r \simeq 16$.  In the 15~orbits of this simulation, V-thin never reaches the point where mixing motions characterized by the sound speed become significant.  The progressively longer times required to approach steady-state illustrate a difficulty with modeling increasingly thin disks: as the transition radius moves out, the ratio of precession rate to orbital frequency diminishes, and the front propagation speed slows down.  Very thin disks require not only greater resolution, but also longer simulations.

\begin{figure}[htbp]
\begin{center}
\includegraphics[width=0.6\textwidth]{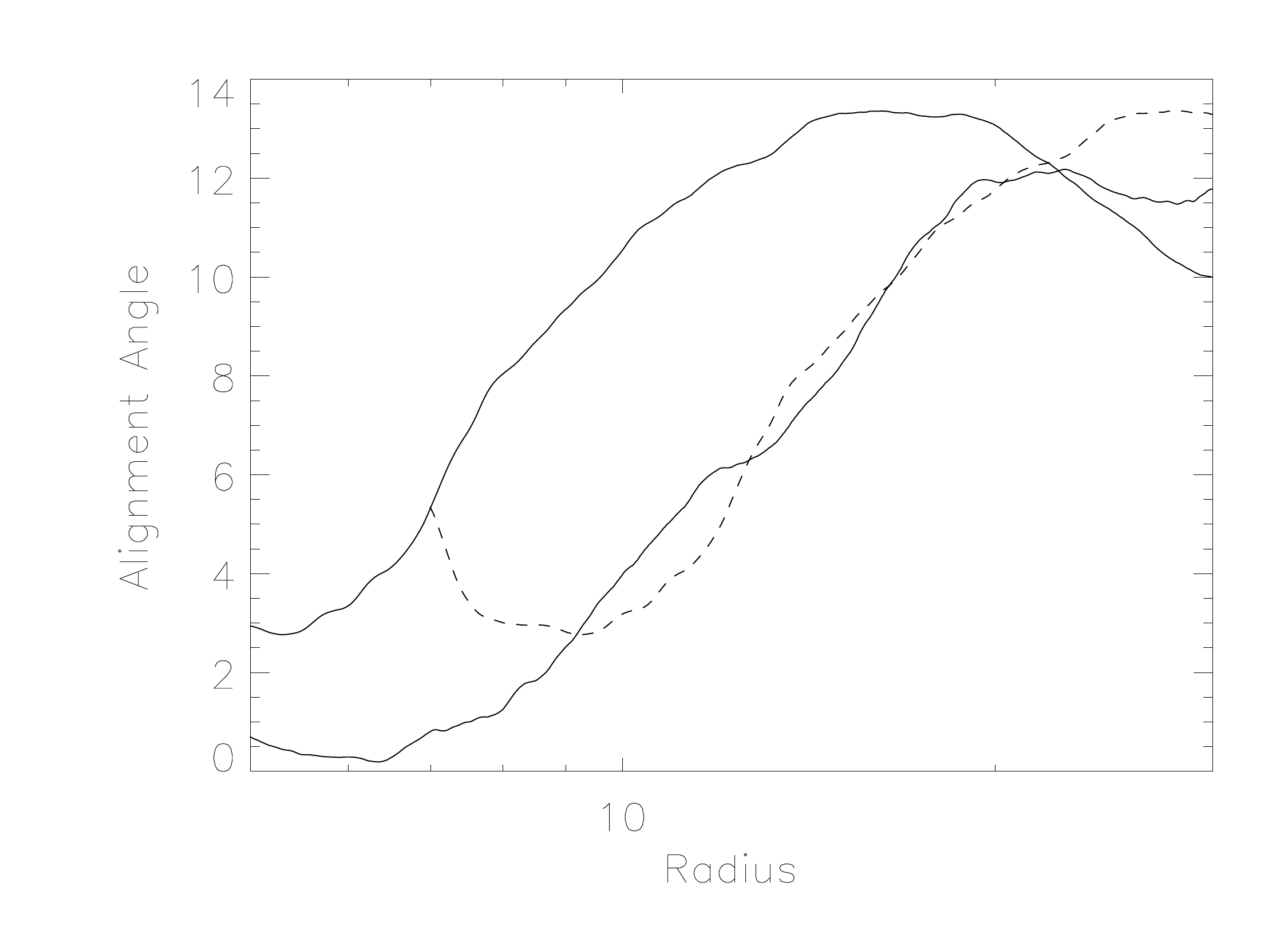}
\caption{Alignment angle $\beta$ as a function of log radius at orbit 10 in KH2015 (left most curve), and at orbit 18 in High-thin (right most solid curve).  The dashed line corresponds to the KH2015 curve moved to the right by a factor of $2^{4/5}$, corresponding to the reduction in sound speed by 2 in going from KH2015 to High-thin. The slope $d\beta/d \log r$ is quite similar for both models.}
\label{fig:beta-vs-r}
\end{center}
\end{figure}

All diffusion models, including ours, predict that the radius of the steady-state alignment front should scale $\propto c_s^{-4/5}$ when the disk is isothermal.  This prediction is tested in Figure~\ref{fig:beta-vs-r}, which shows $\beta(r)$, the alignment angle as a function of radius after alignment steady state has been achieved, at 10 orbits in model KH2015 and 18 orbits in High-thin.  In fact, scaling the radial location of $\beta(r)$ in KH2015 by $2^{4/5}$ (the predicted sound speed scaling) gives excellent agreement not only in the location of the alignment front, but also with respect to its internal structure in the two simulations.   The gradients in alignment across the front in KH2015 and High-thin are nearly identical: $d\sin(\beta)/d\ln r = 0.22$ in the former, 0.20 in the latter.  That the shapes of the curves are so similar when $\beta$ is plotted as a function of $\ln r$ also demonstrates that the ratio of the front's thickness to its radius is nearly independent of $c_s$.

\begin{figure}[htbp]
\begin{center}
\includegraphics[width=0.6\textwidth]{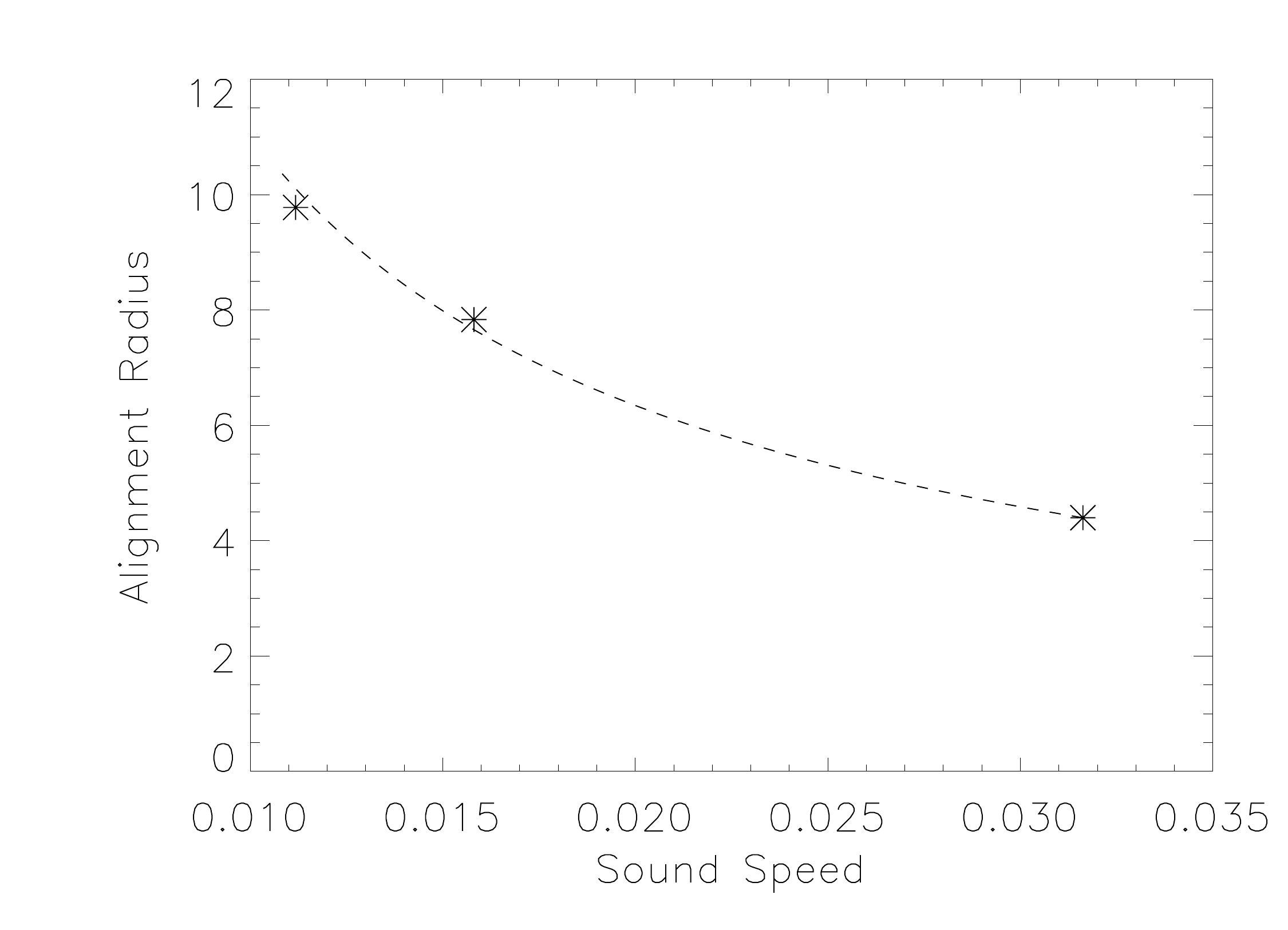}
\caption{The radius where, in steady-state, the alignment angle $\beta = 3^\circ$ plotted as a function of the sound speed for the three runs KH2015 (orbit 12.4), High-thin (orbit 15), and V-thin (orbit 15).  The dashed line shows the location obtained by rescaling the KH2015 point according to $c_s^{-4/5}$.  }
\label{fig:beta-ss}
\end{center}
\end{figure}

V-thin doesn't reach an overall steady state during the time it was evolved, but it does reach a steady state in the inner part of the disk, where the alignment is close to complete.  Figure~\ref{fig:beta-ss} shows the location of the point where $\beta = 3^\circ$ in each of the models as a function of sound speed.  The dashed line corresponds to a rescaling of the KH2015 value $\propto c_s^{-4/5}$.    In addition, the alignment gradient within the front, even though it has not reached its equilibrium location agrees closely with the other two MHD simulations: $d\sin(\beta)/d\ln r = 0.22$ during the final orbit.  Thus, even partial alignment agrees very well with the posited sound speed scaling.

\begin{figure}[htbp]
\begin{center}
\includegraphics[width=0.6\textwidth]{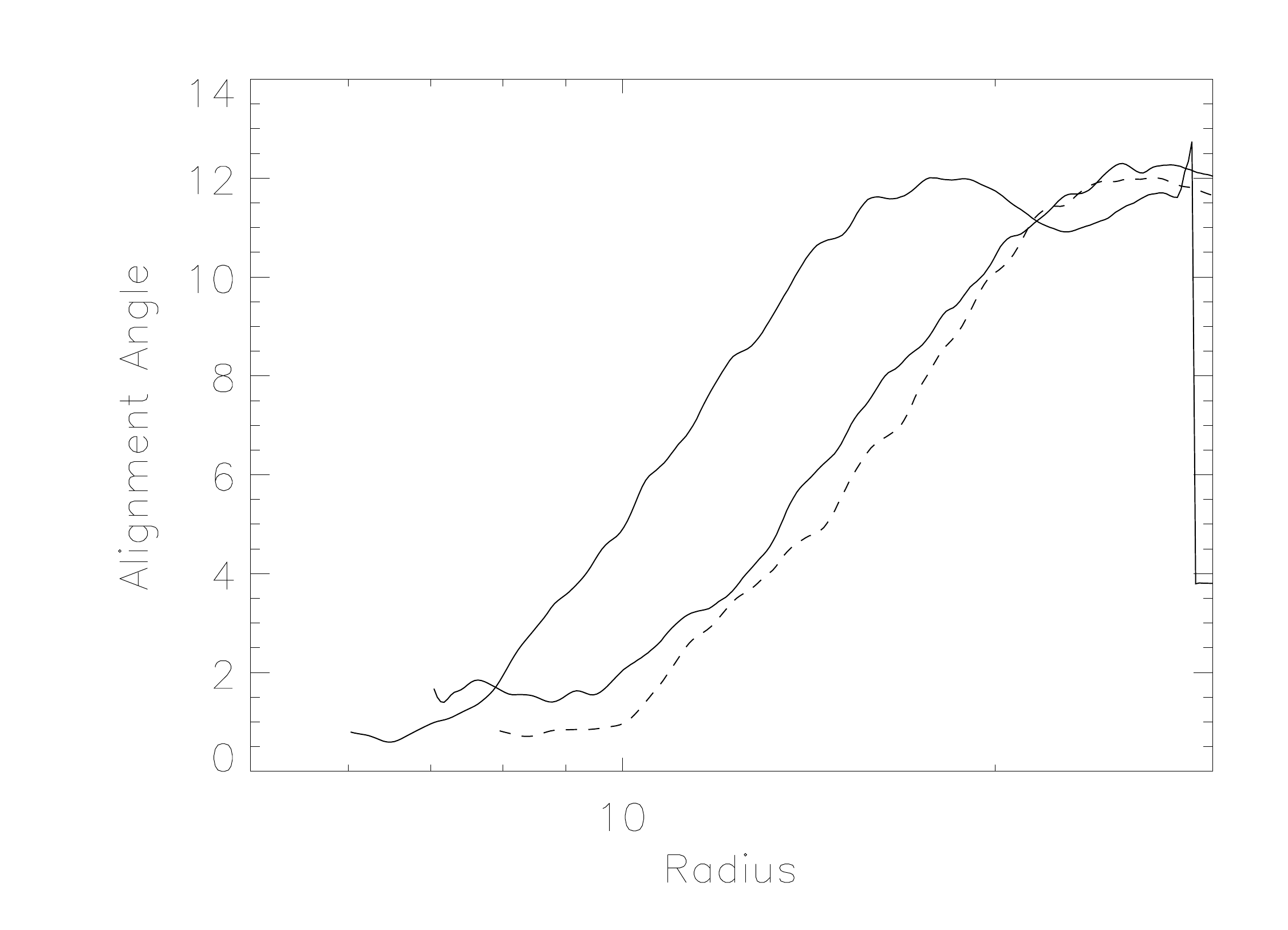}
\caption{Alignment angle $\beta$ as a function of log radius at orbit 15 in Thin-H (left most curve), and V-thin-H (right most solid curve).  The dashed line corresponds to the Thin-H curve moved to the right by a factor of $2^{2/5}$, corresponding to the reduction in sound speed by $\sqrt 2$ in going from Thin-H to V-thin-H.  }
\label{fig:beta-ss-H}
\end{center}
\end{figure}

In fact, the $c_s^{-4/5}$ scaling of the alignment's radial profile applies even more generally than these MHD examples.  A comparison of $\beta(r)$ in two purely hydrodynamic runs, Thin-H and V-thin-H, is given in Figure~\ref{fig:beta-ss-H}.  As in Figure~\ref{fig:beta-vs-r}, the curve for $\beta(r)$ is moved outward by a factor corresponding to the ratio of the sound speeds to the 4/5th power.  The rescaled curve lies nearly upon the curve for V-thin-H, indicating that the sound speed dependence carries over to the HD case as well (although not necessarily the coefficient).   This is not completely surprising, as the model is based on the premise that a hydrodynamic quantity, $c_s$, determines the rate of inward mixing of misaligned angular momentum.   As will be discussed in the following section, there are nevertheless points of both contrast and similarity between paired HD and MHD disks.

The total warp rate $\psi \equiv |\partial \hat \ell/\partial \ln r| = \hat\psi (h/r)$ combines the alignment gradient with the precession phase gradient.   In each of these models, it is spatially noisy within the alignment front, but we can compute mean values at the point where the front stalls.  In KH2015 the radially averaged mean value of $\bar\psi = 0.22$ at orbit 4 between $r=7$ and 12 (corresponding to the locations of $\beta = 6^\circ$ and $10^\circ$.  For High-thin, $\bar\psi = 0.31$ at orbit 12 between $r=12$ and 17.  V-thin does not stall, but at the end of the simulation (orbit 15) $\bar\psi = 0.36$ between $r=4$ and 19.   Contrasting these figures with the alignment gradient, it is apparent that the contribution due to precession phase grows with decreasing sound speed: it is a minor contribution for $h/r=0.1$, but comparable to the alignment gradient's contribution for $h/r=0.035$.

\begin{figure}[h]
\begin{center}
\includegraphics[width=0.5\textwidth]{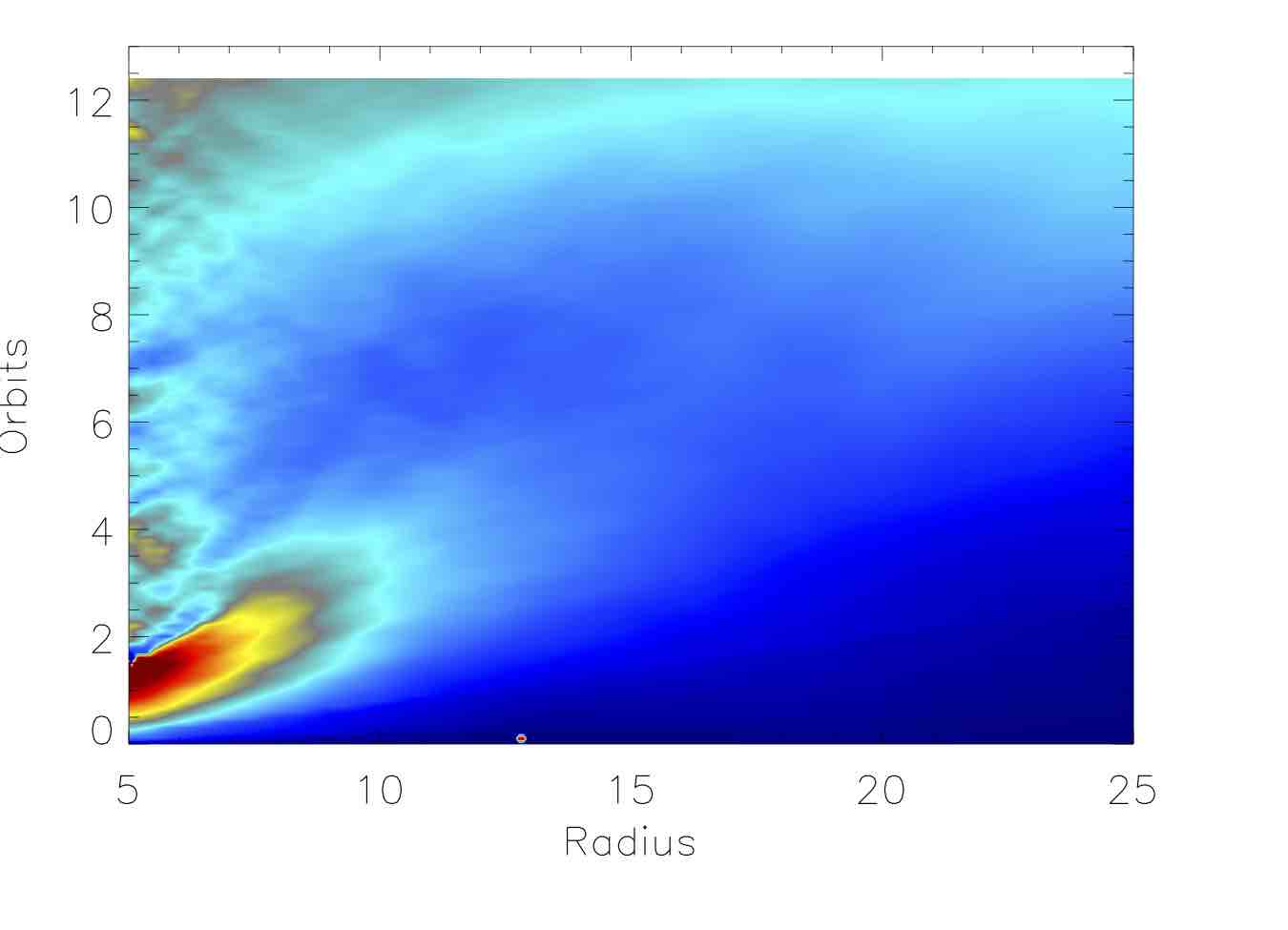}
%\vskip -0.3 truein
\includegraphics[width=0.5\textwidth]{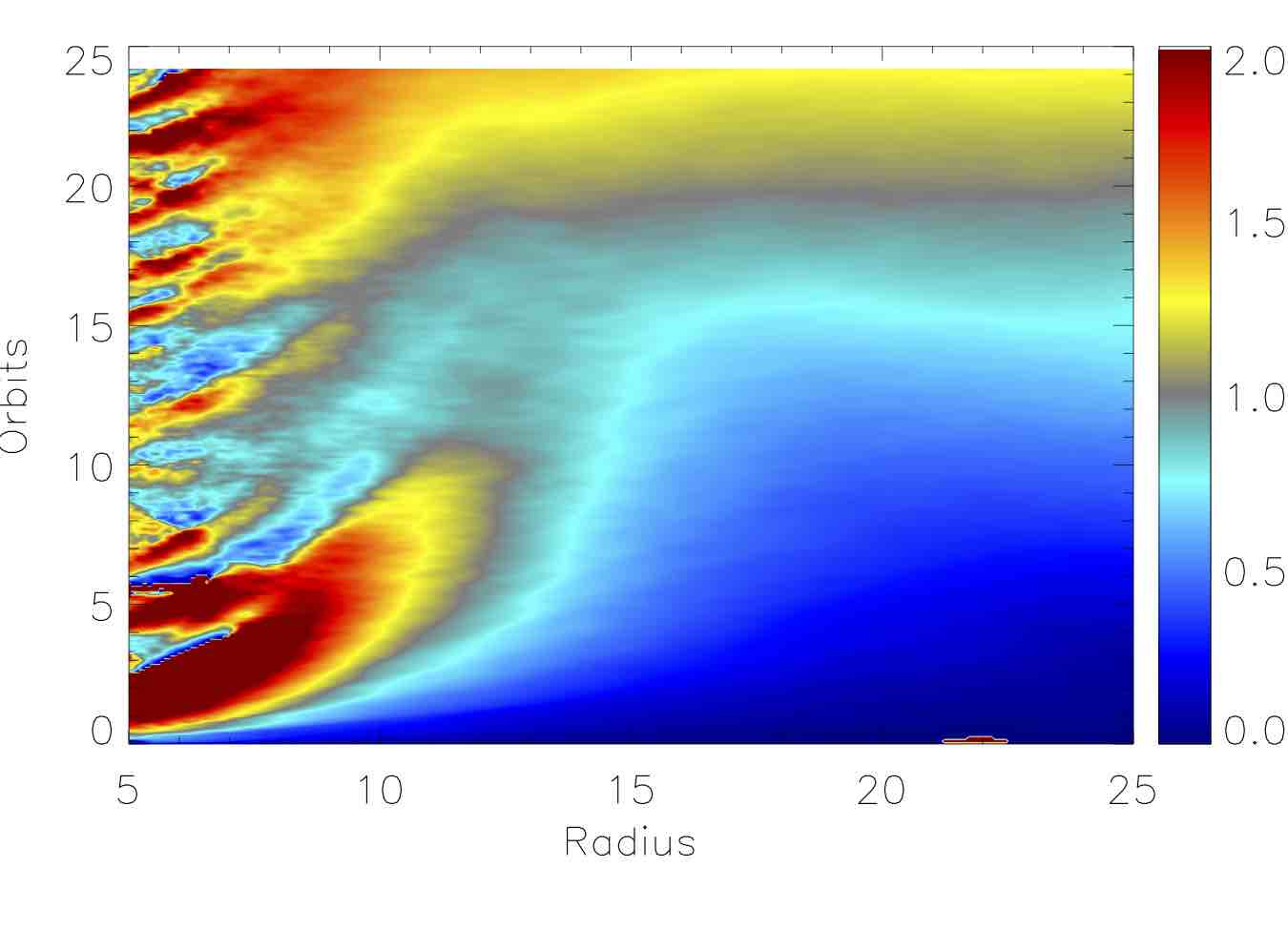}
%\vskip -0.3 truein
\includegraphics[width=0.5\textwidth]{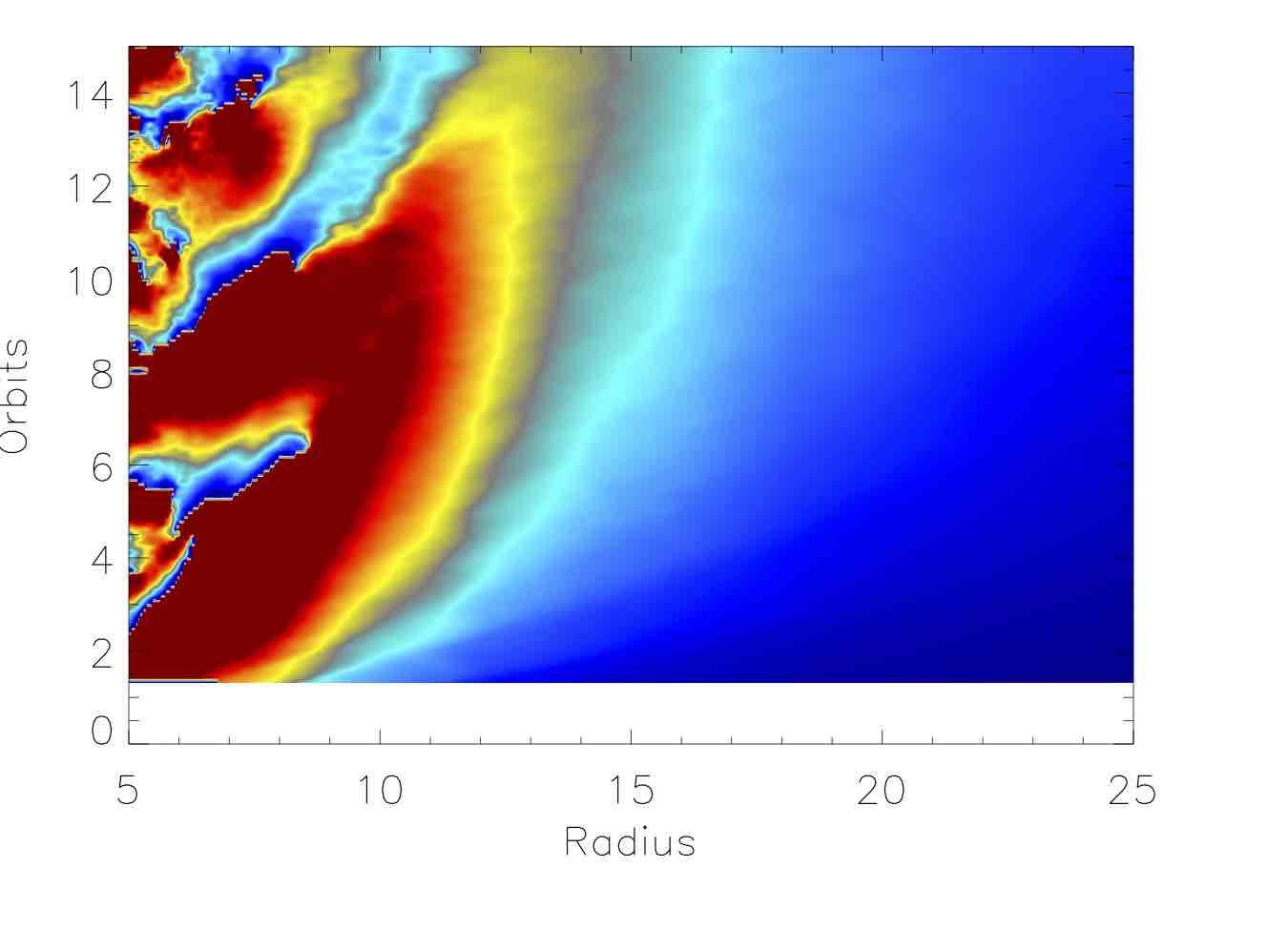}
\caption{Spacetime diagrams for the precession angle $\phi$ in the simulations KH2015 (top), High-thin (middle) and V-thin (bottom).  The angle $\phi$ runs between $0$ and $2\pi$, and the colors run from $\phi=0$ (blue) to $\phi=2$ radians (red) with all $\phi$ angles from 2--$2\pi$ as deep red.  Times of partial solid-body precession are shown by horizontal bands of constant color.  A comparison of the three plots shows that the lower the sound speed, the longer differential precession continues in the disk.  }
\label{fig:phi}
\end{center}
\end{figure}

Figure~\ref{fig:phi} shows spacetime diagrams of the precession angle $\phi$.  Again, there are qualitative similarities shared by all three models. Initially, the disk precesses at the rate given by $\Omega_{precess}$, but soon the precession rate slows and even reverses. Where alignment is almost complete, the precession angle becomes ill-defined, but outside of this region, for example at $r = 7$ in KH2015, $\phi$ stops advancing after 2 orbits, reverses until orbit 3.5, and then advances again at a slower rate thereafter.  Similar behavior is seen farther out in the disk; after $\sim 8$ orbits, the precession phase varies quite slowly with radius outside the alignment front, so that the entire unaligned portion of the disk precesses nearly as a solid-body with a precession frequency close to the LT value at $r=16$\footnote{Although in principle a disk precessing as a solid-body could have a fixed precession phase gradient, i.e., a permanent twist, we use the term as a short-hand for ``uniform precession phase"}.

Differential precession persists for a longer time, and to a greater radius, the lower the sound speed.  Comparison with Figure~\ref{fig:highbeta} shows that, for KH2015 and High-thin, the disk outside the transition front approaches solid-body precession shortly after the transition front reaches its stalling position. In High-thin, precession continues at the Lense-Thirring rate inside $r=15$ until orbit~5, after which precession slows and gradually approaches a rate consistent with $\Omega_{precess}$ at $r=20$. Shortly after orbit 15, the precessional phase becomes nearly independent of radius beyond $r=15$.  By the end of the simulation (22.3~orbits) some radial precessional phase gradient is present from the inner disk out to $r=15$; beyond that point the disk is in solid body precession.  The alignment front in V-thin continues to advance throughout the simulation.  The $r=15$ precession angle $\phi$ advances at the LT rate through orbit~5, after which it continues to advance at a slower rate, but in this case corresponding roughly to the LT rate at $r=25$.  At orbit~15, a radial precessional phase gradient persists from $r=13$ outward. The continued existence of a precession phase gradient over a wide range of radii is consistent with the continuing advancement of the alignment front.

These results are in keeping with previous warped disk simulations, which typically found that solid body precession developed over the entire disk outside the aligned region (if one exists).  \cite{PapTerq95} studied bending waves in a disk induced by the perturbing potential of a binary system.  Viscosity, $\alpha$ or otherwise, was not included.  Their analysis led to an expectation that the disk response could be solid-body precession if its azimuthal sound crossing time is short compared to the precession period.  The numerical simulations of \cite{Fragile05}, for example, found support for this.  Their hydrodynamic (and inviscid) disk, evolved in a relativistic metric,  aligned out to a relatively close-in radius beyond which the disk evolved to near solid-body precession.  This result was interpreted as due to bending waves traveling across the disk faster than the external torque could drive precession.  In our simulations, the time-steady alignment transition is consistently at smaller radius than the point outside which this criterion for bending-wave enforcement of solid-body precession would be satisfied.   In addition, the ability of bending waves to enforce solid-body precession is limited by either passage through turbulence (e.g., driven by the MRI: \citet{SKH13b}) or  shock-damping when the bending wave amplitude is nonlinear (see Sec.~\ref{sec:MHDvsHD}).

Thus, we conclude that, at least over the range of sound speeds considered here, the two central predictions made by the model are supported by simulation data.  The outward speed of the alignment front (when well inside its ultimate stationary location) is  $\simeq 0.35 r \Omega_{precess}$ for the point where 15\% alignment has been achieved, and $\simeq 0.2 r\Omega_{precess}$ for the point where 50\% alignment is achieved, independent of sound speed.  In addition, the steady-state radial location of any given degree of alignment scales $\propto c_s^{-4/5}$, suggesting that, for the purpose of determining steady-state alignment properties, the radial mixing motions can be approximated as a sort of diffusive process.  Moreover, because the second scaling has the corollary that the steady-state position of the alignment front moves outward as $c_s$ declines, while the intrinsic front speed decreases with increasing radius, the time required to reach the steady-state is $\propto c_s^{-12/5}$; colder disks take much longer for the alignment front to reach its steady-state.

\subsection{MHD vs. HD}
\label{sec:MHDvsHD}

\begin{figure}[h]
\begin{center}
\includegraphics[width=0.5\textwidth]{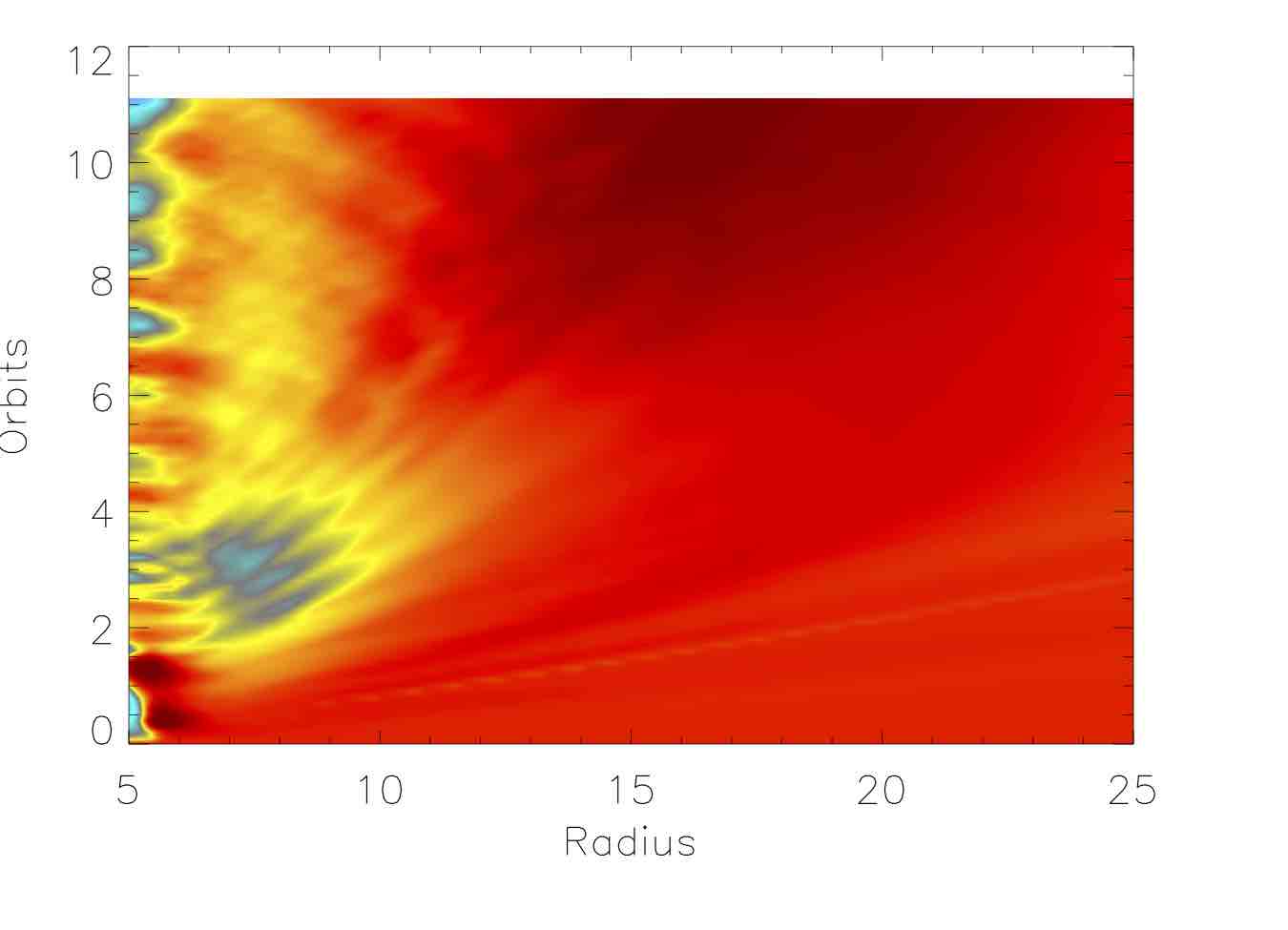}
%\vskip 0.3 truein
\includegraphics[width=0.5\textwidth]{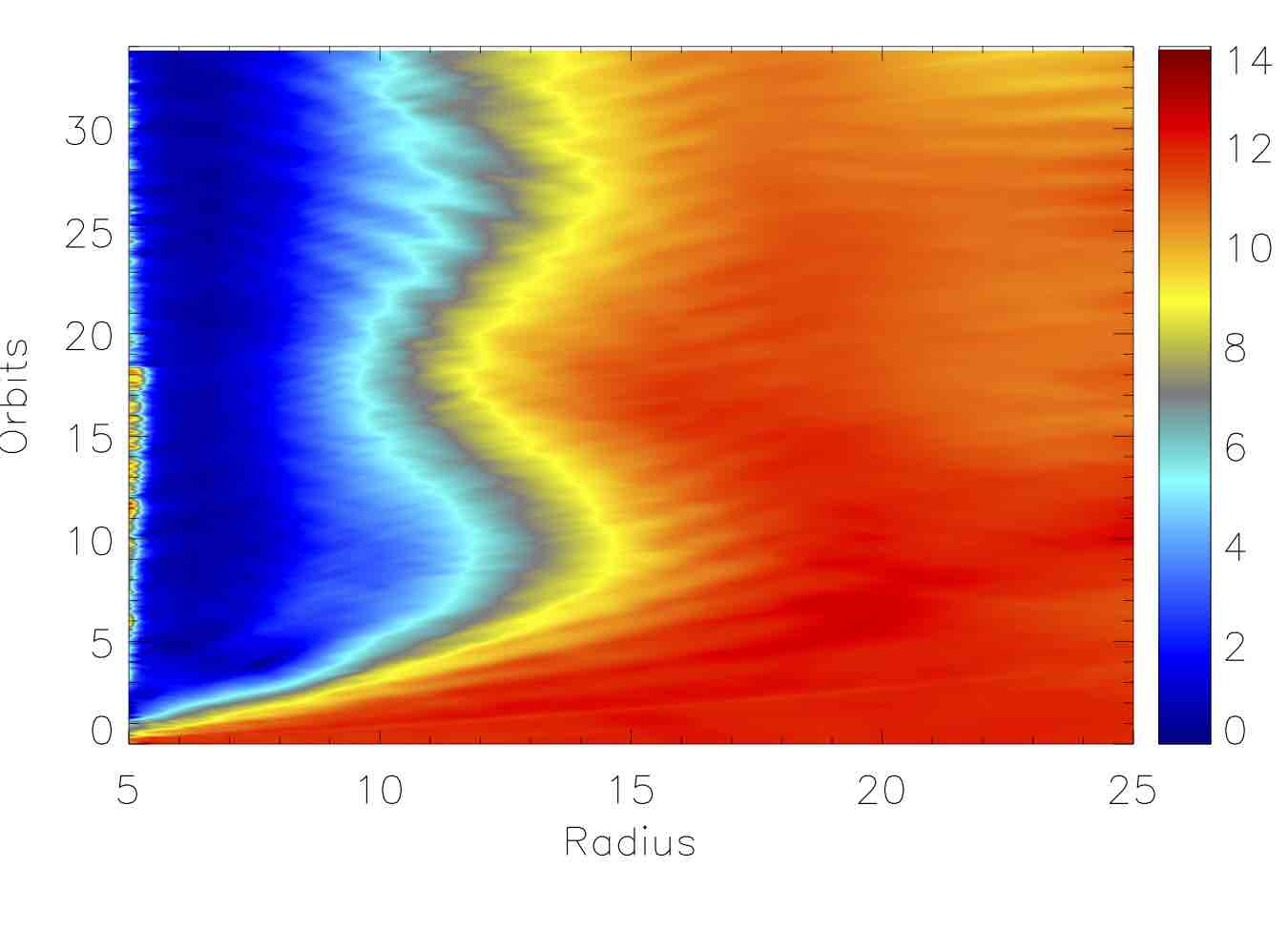}
%\vskip 0.13 truein
\includegraphics[width=0.5\textwidth]{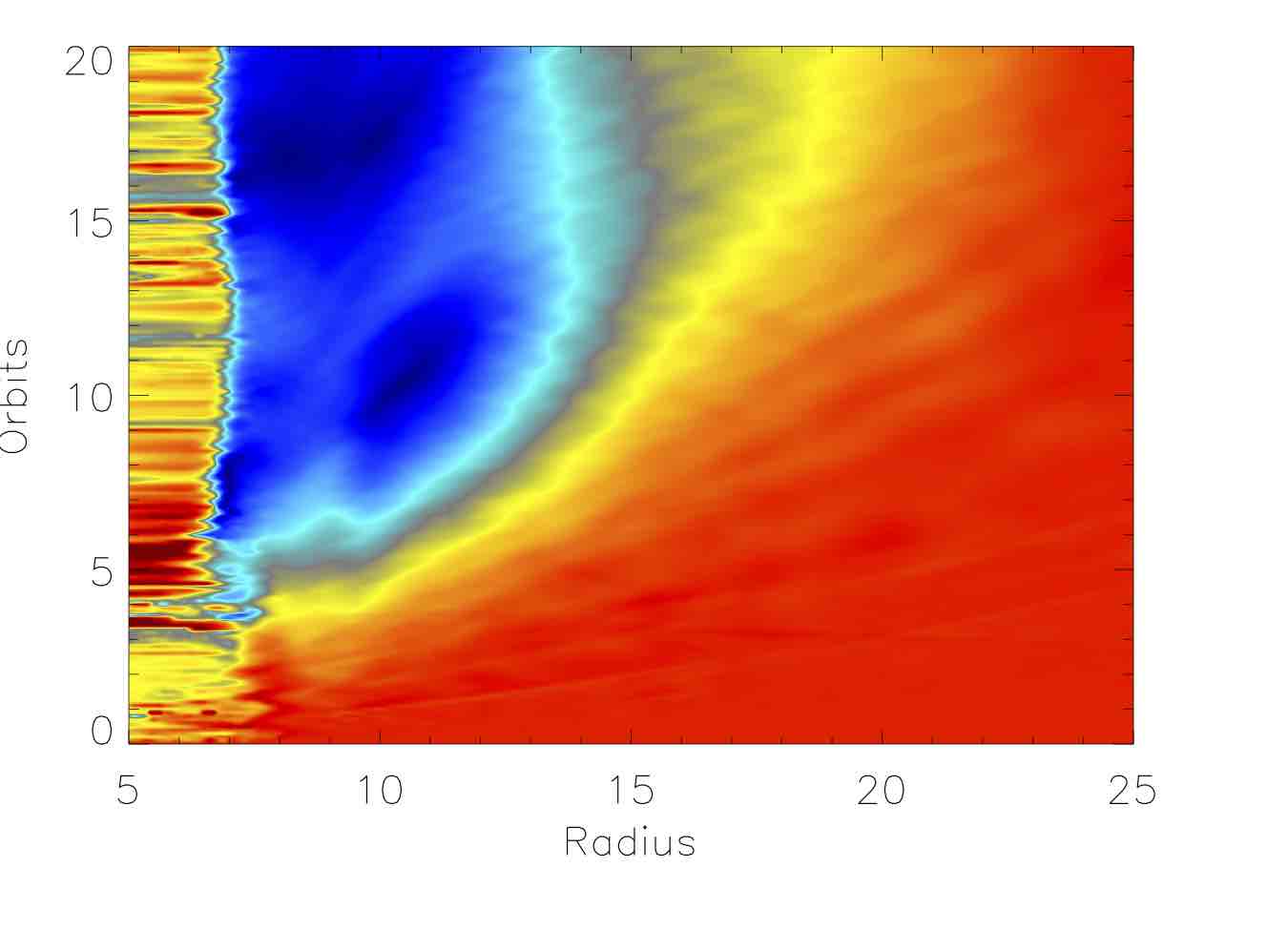}
\caption{Spacetime diagrams for the alignment angle $\beta$ in the  KH2015-H  model (top), Thin-H model (middle) and V-thin-H model (bottom).  Colors run from $\beta=0$ (blue: aligned) to $\beta=14^\circ$ (red).  V-thin-H was evolved from an initial hydrodynamic disk with an inner disk edge at $r=6$. 
}
\label{fig:hydrobeta}
\end{center}
\end{figure}

Figure~\ref{fig:hydrobeta} shows the evolution of the alignment angle $\beta$ in the three hydrodynamic models.  The most striking thing to note is that the hydrodynamic version of KH2015 does not achieve much alignment.  At $r=7$ some alignment occurs in the first 3 orbits, when the inner disk reaches $\beta = 6.6^\circ$, but then it regresses back to $\sim 10^\circ$.
In contrast, the thinner disks show alignment superficially similar to their MHD counterparts.  Thus, the effects of MHD appear to be more dramatic in hotter, geometrically-thicker disks.

Similar behavior was seen by \citet{SKH13b} in their paired HD and MHD simulations.  The disk was relatively thick, with  $h/r \approx 0.12$--0.2 at the beginning of the torqued phase.   Its adiabatic equation of state gave a sound speed that varied in the disk, dropping with radius from $c_s^2 \approx 0.006$ at $r=7$ to 0.001 at $r=20$ (measurements taken at the end of the run).  The inner disk aligned in MHD, but not in HD.  In the HD disks, $\hat\psi \sim 2$--3 in the initial waves that propagate outward following the onset of torque.  After these initial waves move out, $\hat\psi < 1$ throughout most of the disk, and  only partial alignment occurs inside of $r=10$, where $\beta\sim 6^\circ$.  Differential precession ends rather early on; by orbit 5 the disk is near solid-body precession.  \cite{SKH13b} posited that the reason the MHD disk aligned, while HD disk did not, or did so only partially, is that MHD turbulence disrupts the propagation of bending waves, whereas the HD disk remained laminar, permitting bending waves to travel.  This contrast in bending wave behavior is important because bending waves, if allowed to propagate, can quickly lock a wide range of radii in the disk into solid-body rotation, yet a negative radial gradient of precession phase is essential to alignment.  The same explanation also appears to apply to the thinner disk examined by KH2015.  In our HD version of this simulation, KH2015-H, bending waves (easily visible in the top panel of Fig.~\ref{fig:hydrobeta}) travel rapidly outward, enforcing nearly solid-body precession throughout the disk in $\simeq 3$~orbits.  Conversely, in KH2015, bending waves are largely suppressed for the first 5~orbits; the appearance of bending waves traveling outward from the alignment front at that time signals the nearly-simultaneous erasure of precession phase gradients outside the alignment front and the retreat of the alignment front to its long-term location.

Cooler purely HD disks behave differently. Although High-thin (a high-resolution MHD run) and Thin-H (a lower resolution HD run) show different histories of orientation front overshoots and retreats, by the time 20--30~orbits have passed, both runs have reached near steady-state orientation profiles, with the front located at very nearly the same place (contrast the middle panels of Fig.~\ref{fig:highbeta} and Fig.~\ref{fig:hydrobeta}; note the different durations of the runs).  Thus, in contrast to KH2015-H, a purely HD evolution is able to align in a way that ultimately resembles an MHD disk.  Interestingly, the alignment behavior of Low-thin (a low-resolution MHD run) is intermediate between High-thin and Thin-H: the greater magnetic diffusivity created by poor resolution makes Low-thin an only weakly MHD simulation.  The still cooler pure HD run V-thin-H behaves similarly to Thin-H in that the initial alignment front stalls and partially reverses by orbit 15, while the alignment front in the  MHD version continues to move outward.

\begin{figure}[h]
\begin{center}
\includegraphics[width=0.5\textwidth]{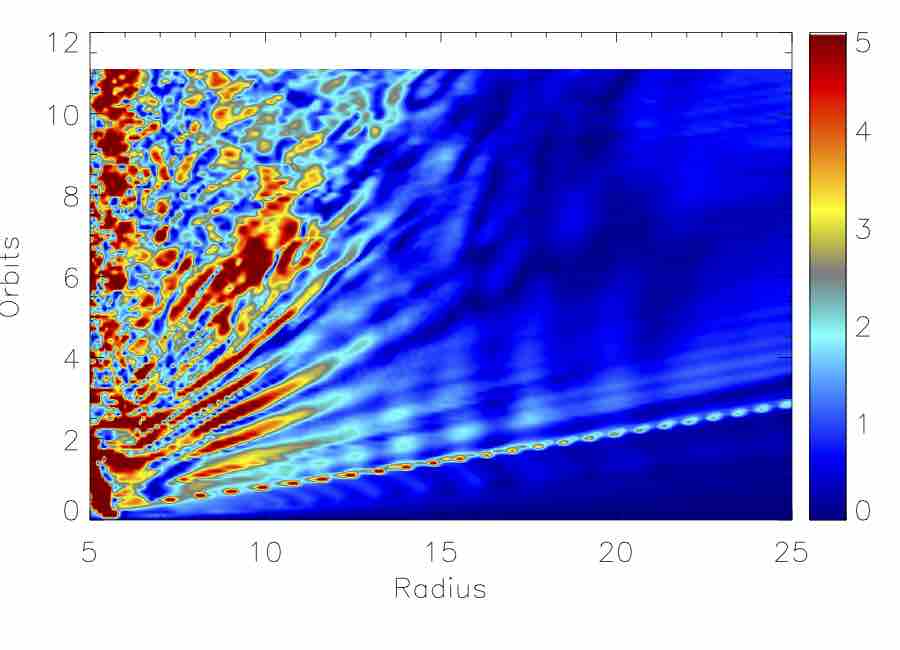}
%\vskip -0.3 truein
\includegraphics[width=0.5\textwidth]{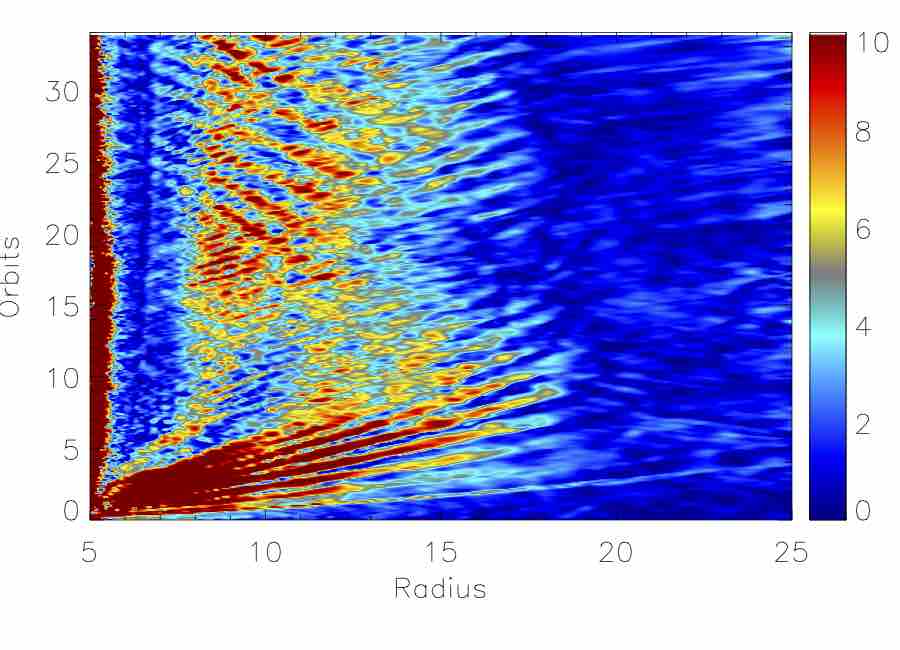}
%\vskip -0.3 truein
\includegraphics[width=0.5\textwidth]{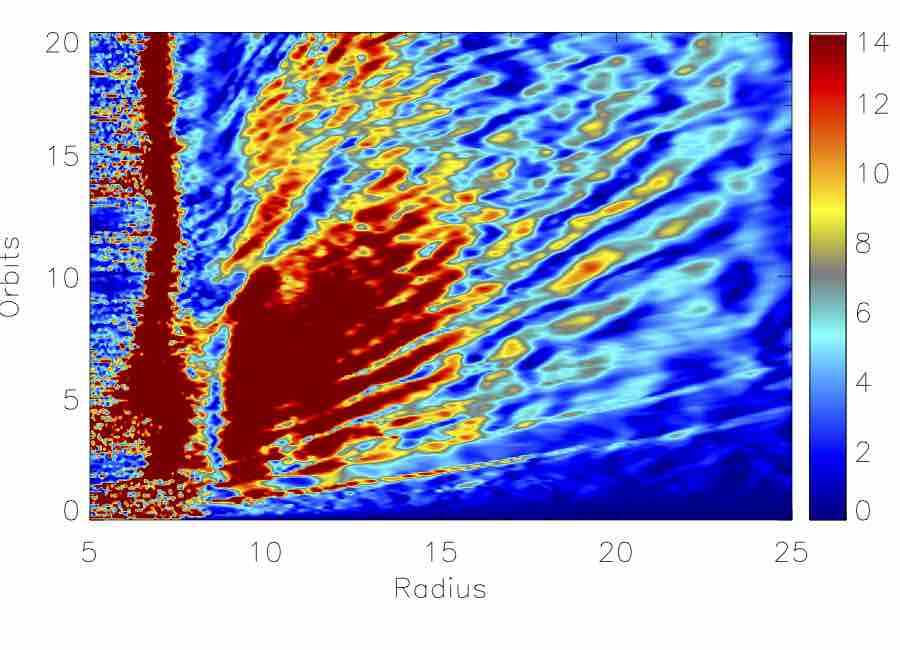}
\caption{Spacetime diagrams for the normalized warp $\hat\psi$ in simulations KH2015-H (top), Thin-H (middle), and V-Thin-H (bottom).  Each plot has a its own color scale, so that the relative scales are proportional to the relative sound speed in each model.  From this it is clear that $\hat\psi$ is larger in colder disks by an amount that is greater than that which can be attributed to the decrease in $h/r$.   V-thin-H was evolved from an initial hydrodynamic disk with an inner disk edge at $r=6$. }
\label{fig:hydropsihat}
\end{center}
\end{figure}

To understand why cooler hydrodynamic disks are better able to align, we look more closely at the character of the bending waves as a function of sound speed.  Just as in KH2015-H, persistent inward- and outward-traveling bending waves are visible in both of the thinner HD disks throughout their evolutions, although the inward-directed waves are considerably weaker in V-thin-H.  However, KH2015-H differs significantly from the others in its normalized warp amplitude.  The typical unnormalized warp amplitude of the initial bending waves in KH2015-H is $\psi \sim 0.26$, but the fiducial value of $h/r$ in KH2015-H, $\approx 0.1$, making the initial $\hat\psi \sim 2$--3. By contrast, in both Thin-H and V-thin-H, $\hat \psi$ is larger since $h/r$ is smaller (at the fiducial radius, it is $\sim 0.05$ and $\sim 0.035$, respectively).  Consequently, the normalized warp $\hat\psi$ increases with decreasing sound speed,  making the cooler disks' bending waves increasingly nonlinear in the initial orbits after the torque is turned on (first $\simeq 5$~orbits for Thin-H, first $\simeq 10$--12~orbits for V-thin-H).  The spacetime diagram for $\hat\psi$ is shown in  Figure~\ref{fig:hydropsihat}.  In each of the three figures the color scale is proportional to the sound speed.  The increasing prevalence of red as the sound speed decreases shows that  $\hat\psi$  increases by a greater factor with decreasing sound speed than can be attributed solely to the reduction in scale height.  Note too that since $h/r \propto r^{1/2}$ in these isothermal simulations, $\hat\psi$ declines with radius, all other factors being equal.  In addition, when the bending waves begin at a nonlinear amplitude, $\psi$ diminishes as they travel, making the decline of $\hat\psi$ more rapid than $r^{-1/2}$.

As shown by \citet{SKH13a}, the propagation of linear and nonlinear bending waves is quite different.  The former can propagate long distances (in laminar disks) with little damping, while the latter rapidly diminish in amplitude through shock formation; indeed this diminution can be seen along the wave tracks displayed in Figure~\ref{fig:hydropsihat}.  In Thin-H and V-thin-H, this diminution in amplitude renders the waves less effective in enforcing solid-body rotation than they are in KH2015-H.  Greater persistence of radial precession phase contrast follows, and alignment can continue so long as negative radial precession phase gradients endure.  By this means, as shown in Figure~\ref{fig:hydrobeta}, the two thinner HD disks create sizable aligned regions.

\begin{figure}[h]
\begin{center}
\includegraphics[width=0.5\textwidth]{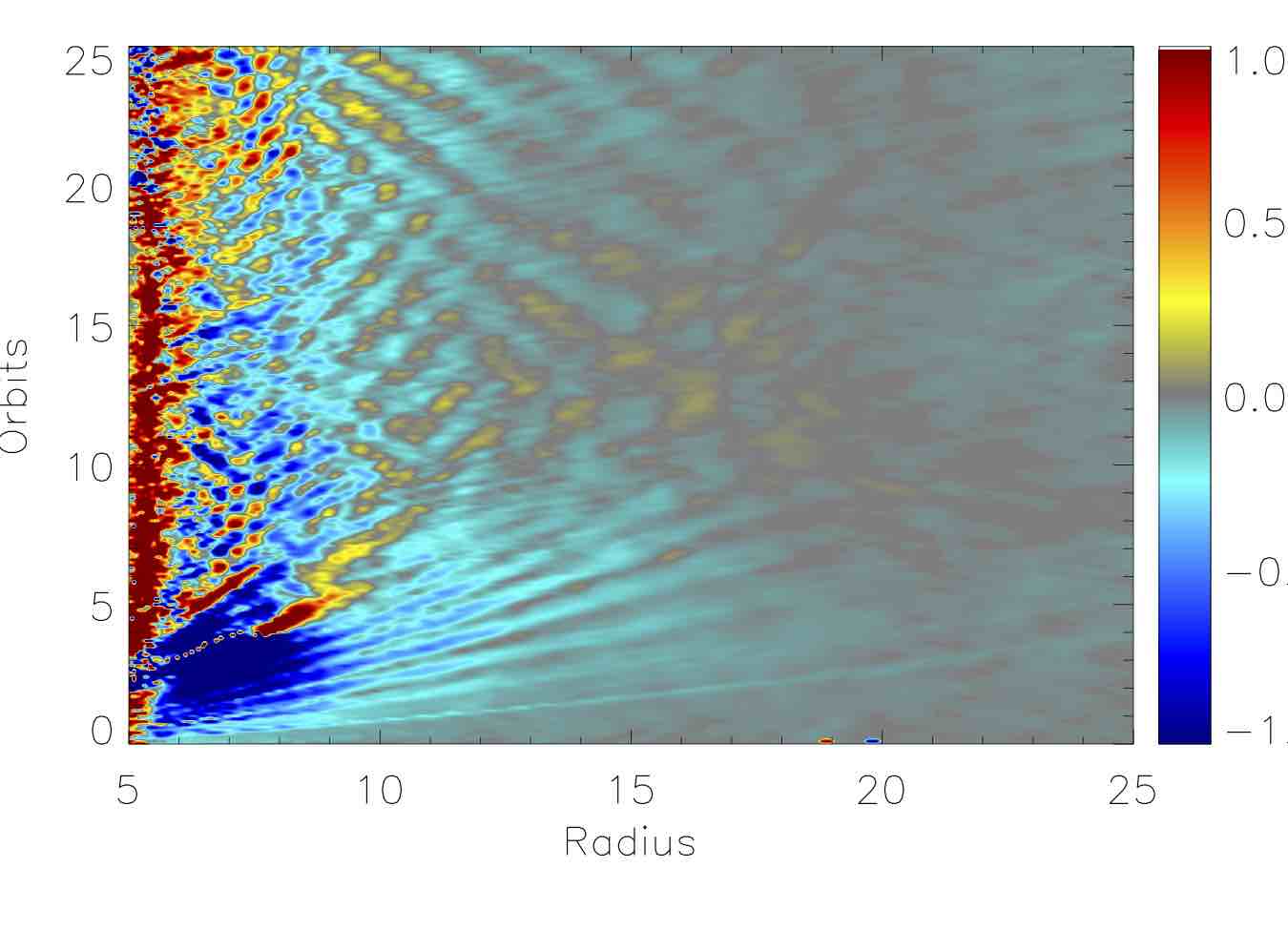}
%\vskip -0.3 truein
\includegraphics[width=0.5\textwidth]{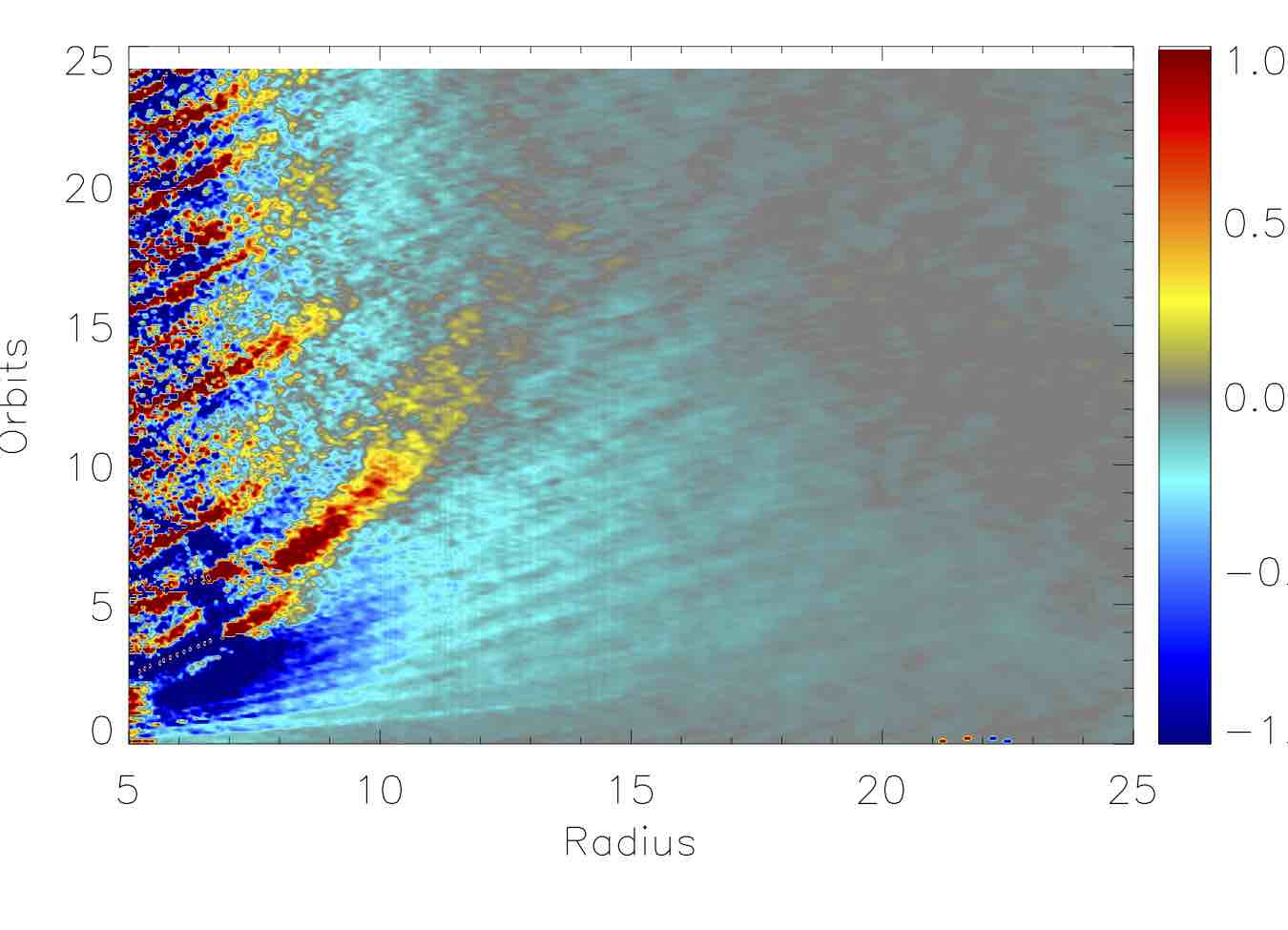}
\includegraphics[width=0.5\textwidth]{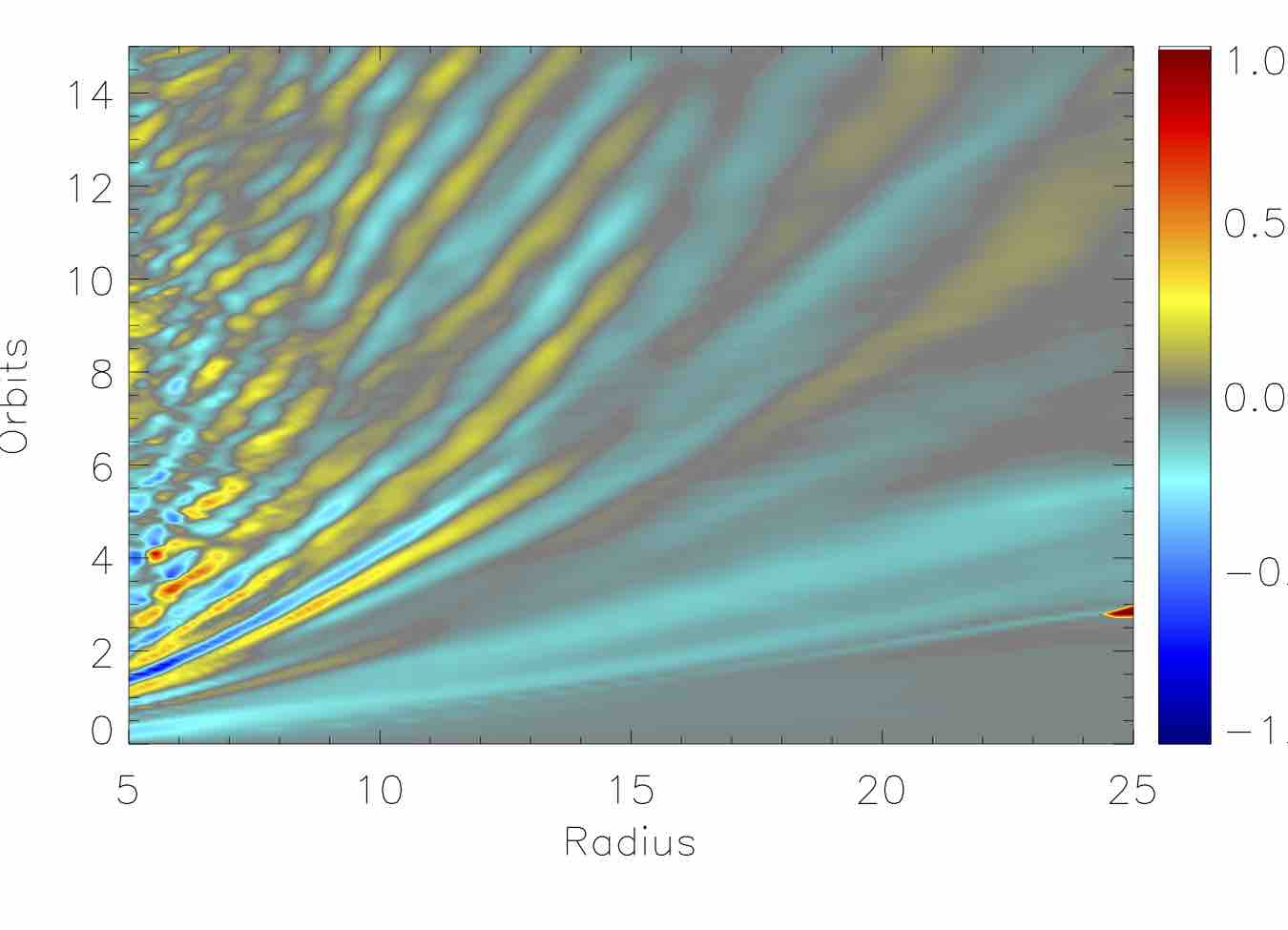}
\caption{Spacetime diagrams for the radial precession phase gradient $d\phi/dr$  in simulations Thin-H (top), and High-Thin (middle), and the HD simulation from \cite{SKH13b}.  }
\label{fig:dphidr}
\end{center}
\end{figure}

Figure~\ref{fig:dphidr} is a spacetime diagram of the radial precession phase gradient, $d\phi/dr$.  In this figure the blue and cyan colors correspond to negative gradients, a requirement for alignment.  Regions that are grey have no gradient, while yellow-red regions have the wrong sign for alignment.  In the top figure, for Thin-H, the systematic cyan color in the region $r>10$ gives way to an unfavorable gradient after about 10 orbits.  This corresponds to the reversal of the alignment front after orbit 10 as seen in Figure~\ref{fig:hydrobeta}.    The criss-cross pattern of waves through the disk is visible in the disk after 10 orbits, with both positive and negative gradients; the average is, however, slightly negative and the alignment front recovers and stabilizes.  The MHD model (middle figure) is  smoother, and the region of negative gradient persists for a longer time, until orbit 15.  The ``finger" of large positive $d\phi/dr$ running from
$(r=8,t=6)$ to $(r=10,t=9)$  lies along the edge of a nearly-flat aligned region which is no longer precessing; the angle $\phi$ ``jumps'' back to the value given by $\Omega_{precess} t$.  Finally, the HD simulation from \cite{SKH13b} provides an example of the near absence of a precession phase gradient when $\hat\psi$ is in the linear regime.  Following a brief initial period the disk simply experiences outward moving waves with no net gradient as the disk precesses as a solid-body  \citep[see Fig.~3 in][]{SKH13b}. 

The basic criterion of the ability to maintain precession phase gradients controls the effectiveness of alignment in both MHD and purely HD disks.   Because, for fixed external torques, cooler disks have more nonlinear warps, and are therefore less effective in creating regions of solid-body rotation, cooler disks are also more able to align.  This, too, is illustrated in Figure~\ref{fig:hydrobeta}: the aligned region in V-thin-H extends to larger radii than in Thin-H.  Similarly, the realignment episode seen between orbits 18 and 26 in Thin-H is associated with the reappearance of a precession phase gradient near the alignment front, even though the outer regions of the disk at that time exhibit very nearly solid-body precession. 

\begin{figure}[h]
\begin{center}
\includegraphics[width=0.5\textwidth]{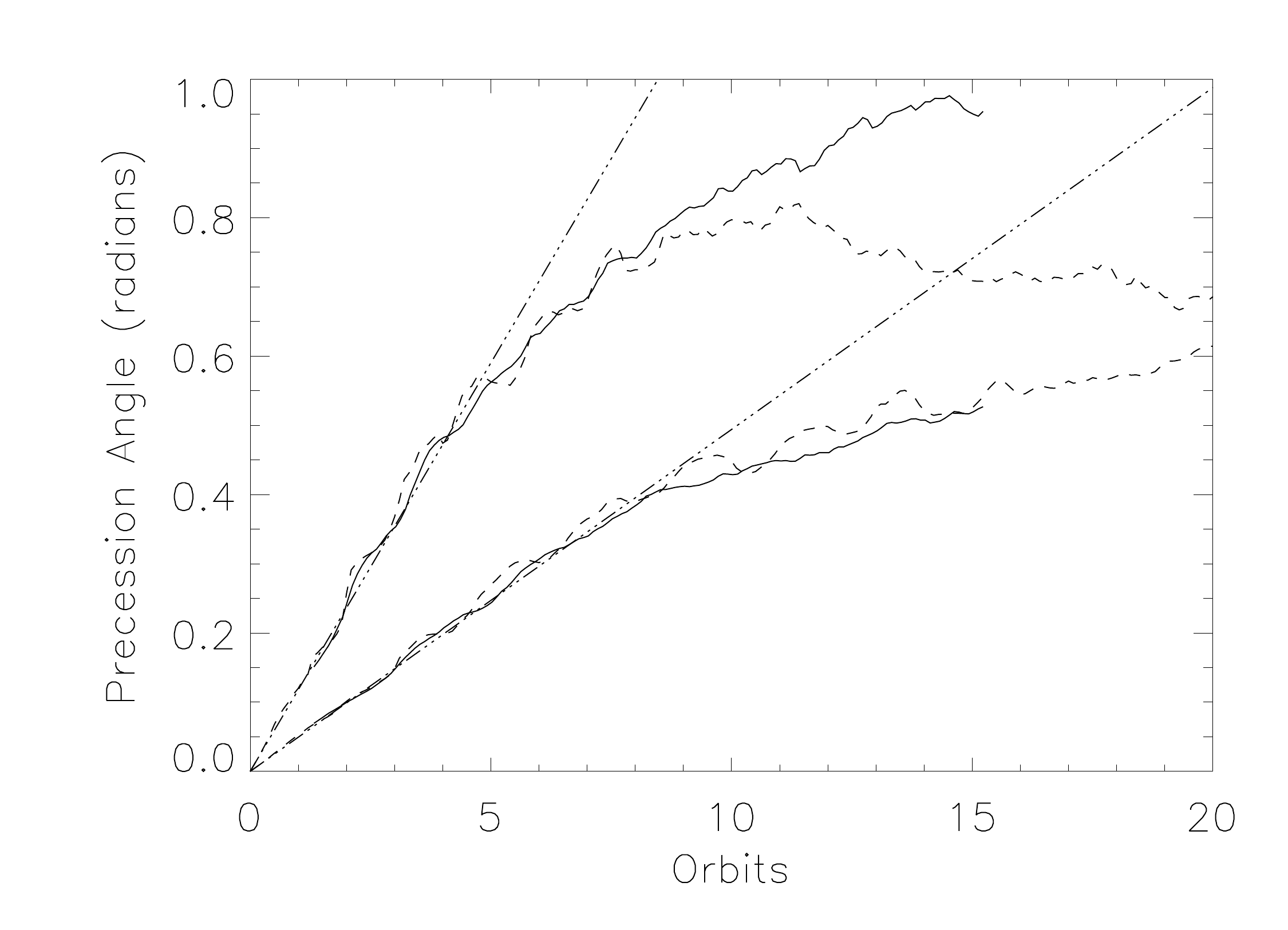}
\caption{Evolution of the precession angle $\phi$ as a function of time at $r=15$ (upper pair of curves) and $r=20$ (lower pair of curves) for V-thin (solid line) and V-thin-H (dashed line).  The straight lines (dot-dashed) show the value of $\phi$ given by $\Omega_{precess}t$ at the selected radii.
}
\label{fig:V-thin-phi}
\end{center}
\end{figure}

In fact, V-thin-H succeeds so well with alignment that in many ways it resembles its MHD partner V-thin. In V-thin-H the alignment front moves at roughly the same pace as it does in V-thin, and by orbit 15 the alignment front has arrived at very nearly the same location in both.  This early similarity in alignment is mirrored in similarity of their precession phases (Fig.~\ref{fig:V-thin-phi}).  However, this apparent elimination of contrast between HD and MHD disks may be a symptom of the slowness with which low sound speed disks' alignment properties evolve.  Contrasting their alignment histories (Fig.~\ref{fig:highbeta} vs. Fig.~\ref{fig:hydrobeta}), it is apparent that whereas the alignment front in V-thin moves out monotonically, the alignment front in V-thin-H began to move inward at $t \simeq 15$~orbits.  Consideration of their distribution of precession phases at this time suggests that at later times the degree of alignment in these two simulations will diverge.  At the latest time in V-thin (15~orbits), the precession phase continues to decline outside the alignment front, while in the last $\simeq 6$~orbits of V-thin-H ($t = 15$ to $t=21$), the magnitude of the precession phase gradient steadily declines, so that by $t=20$, there is almost no phase contrast from $r=15$ to $r=20$ (Fig.~\ref{fig:V-thin-phi}).  If our analysis of alignment dynamics is correct, the alignment front in V-thin will continue to move outward, whereas that in V-thin-H has already begun to turn around and move to smaller radii.

Thus, we arrive at the conclusion that when HD disks are cool enough for their bending waves to be strongly nonlinear at their launch points, they are able to achieve steady-state alignment configurations very similar to MHD disks.  However, the process by which they arrive at that steady state can be somewhat different, especially in the sense that MHD disks tend to overshoot the ultimate steady-state alignment location by larger amounts than HD disks do.

\subsection{Very large disks}

As we have already remarked, inward-traveling as well as outward-traveling bending waves can be clearly seen in Figure~\ref{fig:hydrobeta} in Thin-H, and more weakly in V-thin-H.  Their presence raises the question as to whether waves reflecting off the outer boundary of the disk are overly influencing the outcome.  To test this hypothesis, we ran a  model (``Big-H'') with the same temperature as in Thin-H, but whose outer disk boundary was moved to $r=100$ so that no waves reflected from there could reach the inner disk during the simulation.  In this model, the alignment front starts off much as in V-thin-H, with the head of the alignment front (defined here as the radius where $\beta = 10^\circ$) reaching a radius comparable to that achieved in High-thin, $r=17$, at $\simeq 10$~orbits.  However, unlike High-thin, in Big-H the front then retreats abruptly to $r=13$, similar to the location of the late-time front in Thin-H, and stays near there until the end of the simulation at $\simeq 22$~orbits.  Consistent with this behavior, the disk outside the alignment front has essentially no remaining precession phase contrast after $\simeq 10$~orbits.  In this sense, the outer radius of the disk appears to make little difference.

\begin{figure}[h]
\begin{center}
\includegraphics[width=0.5\textwidth]{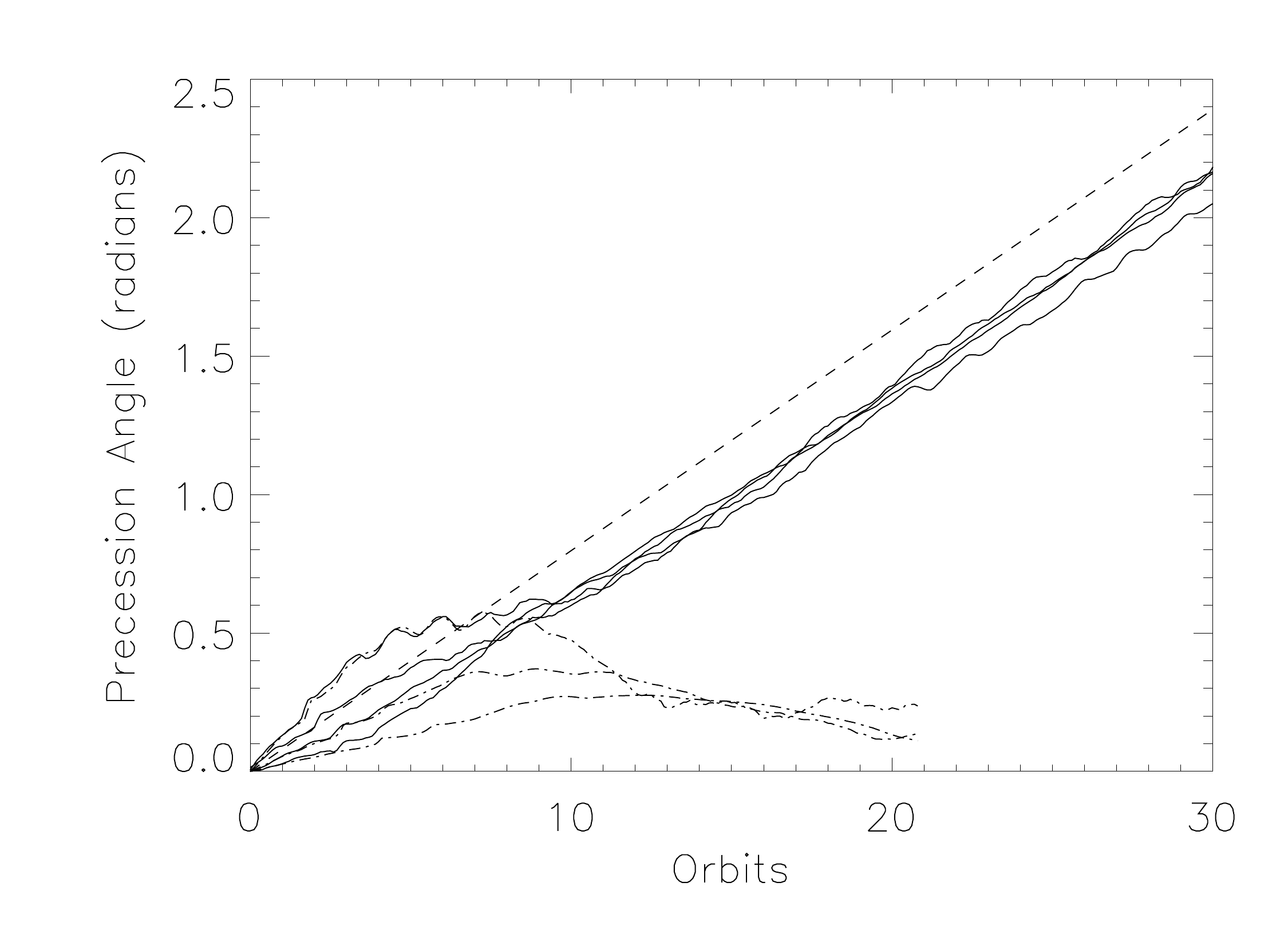}
\caption{Evolution of precession angle $\phi$ as a function of time at specific radii in two hydrodynamic disk models.  The solid lines are from Thin-H at locations $r=15$, 17, 20 and 25.  The dot-dashed lines are from Big-H at locations $r=15$, 20, and 25.  The dashed line shows the angle corresponding to precession at the LT rate at $r=17$, the rate at which the outer disk in Low-thin evolves toward at late time.  Big-H, on the other hand, evolves toward a very low, or zero, precession rate in the outer disk.}
\label{fig:Big-H-phi}
\end{center}
\end{figure}

Larger disks do, however, affect the precession rate of the solid-body precession in the outer disk (see Figure~\ref{fig:Big-H-phi}).  In Thin-H, the region $r>15$ precesses at a rate equal to the LT rate at $r=17$.  By contrast, in Big-H, the disk at $r > 15$ briefly precesses {\it retrograde} from $\simeq 9$--12~orbits, and then hardly precesses at all for the rest of the simulation.  Such slow solid-body precession corresponds to the rate at much larger radius, a situation consistent with the fact that both the mass and the angular momentum of Big-H are dominated by contributions at very large radius: $\Sigma$ peaks at $r \approx 55$.  Thus, while the size of the disk seems to have minimal impact on the inner disk alignment, it has a substantial impact on the late-time solid-body precession rate in the unaligned disk outside of the alignment region.  
 
\section{Discussion: the operation and regulation of radial mixing}

An understanding of alignment processes in torqued disks rests on four basic ideas: (1) the new angular momentum required to change disk orientation can come only from the external torque; (2) angular momentum delivered at an inner radius is carried outward by radial fluid motions induced by the disk warp; (3) the angular momentum delivered by the external torque can contribute to alignment only if the precession phase decreases outward,  hence the on-set of solid body precession can end alignment; and (4) disk orientation ceases to change when similarly warp-induced fluid motions convey misaligned angular momentum inward at a rate matching the outward transmission of aligning angular momentum.  This picture implies that when the outward transmission of aligning angular momentum dominates, the location in the disk where its alignment changes from aligned to misaligned moves outward at a rate $\sim r \Omega_{\rm precess}$.  Because the radial fluid motions are generically transonic when the disk warp is nonlinear \citep{SKH13a}, this picture implies that the balance between growth and diminution of the aligned region depends on sound speed; the aim of this paper has been to test the predicted dependence with numerical simulations.

Because simulations of this problem are computationally expensive, we have investigated this dependence in the sense of measuring a partial derivative: we have changed the sound speed without altering anything else.  Moreover, to simplify what is meant by ``the sound speed", we have focused on studying isothermal disks.  We have also restricted our attention to a single surface density profile ($\Sigma \propto r^{3/2}$,  except where the disk is truncated at its boundaries), one that does not correspond to inflow equilibrium for this temperature distribution.  In fact, even within this model, we have been able to sample only three values of the sound speed: measured in units of the orbital speed at our fiducial radius, these values are 0.1, 0.05, and 0.035.  Nonetheless, within these constraints, we have found three notable results:

\begin{itemize}

\item Within the alignment front  the coefficient in the expression for the alignment front speed is $\simeq 0.35$ at the point at which the initial disk obliquity has been reduced by $\simeq 15$\%, and $\simeq 0.2$ where the disk obliquity has been reduced by half,  independent of sound speed and independent of whether the disk is MHD and turbulent or HD and laminar; however, for laminar HD disks to align at all, they must be sufficiently cool that bending waves are launched at nonlinear amplitude.

\item Approximating the radial flow of angular momentum, both inward and outward, as a diffusive process acting on an alignment gradient of fixed $|\partial \hat\ell /\partial \ln r|$ leads to the prediction that the alignment front becomes stationary at a radius $\propto c_s^{-4/5}$; this prediction is supported very strongly across the range of sound speeds examined.

\item Maintenance of a precession phase gradient against the tendency of angular momentum diffusion to enforce solid-body precession requires disruption of bending wave propagation.  This is most effectively accomplished by fluid turbulence, which, in this context, is stirred by the magnetorotational instability, but it can also be accomplished, when disks are sufficiently cool, by the shock-damping of the fluid's radial motions that occurs when the disk warp is nonlinear, i.e., $|\partial \hat\ell /\partial \ln r | > h/r = c_s/v_{\rm orb}$.

\end{itemize}

In this section we expand upon these points and present some further consequences of these results.

\subsection{The diffusion model of radial mixing}

We begin with deriving the dimensionless factor $A$ in the orientation diffusion coefficient.    Solving Eq.~\ref{eq:rT}, we find
\begin{eqnarray}
A &=& B(r_T)^{-1} \frac{\Omega_{\rm precess}(r_*)}{\Omega(r_*)} \langle \cos \gamma\rangle {\cal I} (r/h)^2 (r_T/r_*)^{-5/2}\\
   &=& 0.009 B(r_T)^{-1} (r/h)^2 (r_T/r_*)^{-5/2}.
\end{eqnarray}
Data from our simulations permits evaluation of all the relevant quantities.   Using the measured values of $d\sin \beta/d\ln r$ for KH2015 and High-thin to calculate $B(r_T)$, and having found ${\cal I} \simeq 0.3$ for both of these simulations, the steady-state alignment front locations for KH2015 and High-thin then imply $A=2$ and $A=1.5$, respectively.   In other words, these results suggest that the dimensionless factor multiplying the dimensional form $c_s^2/\Omega$ is close to a constant $\simeq 2$.

It is surprising how well the simple diffusion approximation works, over a factor of 8 in temperature, in predicting time-steady properties of both our turbulent MHD and inviscid HD models.  The diffusion approximation posits that the flux of some quantity is proportional to the gradient in its density, but \citet{SKH13a} showed that the angular momentum flux induced by a given level of disk warp is both delayed with respect to the creation of that warp (by a time comparable to an orbital period) and dependent upon the radial range over which the gradient extends, not just the gradient's local value.  In fact, our simulations already show that the position of the alignment front generically oscillates about its ultimate time-steady location, behavior inconsistent with a diffusion model.

Although we cannot state with confidence why diffusion fails to describe the time-dependence of warp evolution, yet succeeds as a guide to the time-steady solution, there are some plausible arguments that could provide a partial explanation.   The delay between angular momentum flux and warp creation becomes less and less relevant as conditions approach steady-state.  The simple sound speed scaling may be a product of all our simulations sharing the same surface density profile and tilt angle, as well as possessing a single sound speed everywhere.  As a result, the several shape factors are nearly the same, both $|\partial \hat\ell /\partial \ln r|$ across the alignment front and the dimensionless integral ${\cal I}$ (eq.~\ref{eq:scrI}).  In addition, because the dynamical factors relevant to this problem (the Newtonian orbital frequency, the Lense-Thirring torque) are both power-laws in radius, and therefore scale-free, and because all the simulations shared the same scale height and surface density profiles as well as the same intrinsic misalignment angle, it was possible for all these shape factors to remain invariant to changing sound speed.

It is also important to point out that much previous work has been done casting warp dynamics in terms of diffusion models.  \cite{NP00}, for example, discussed several potential formul{\ae} for the transition radius, all of them based on diffusion models, but with diffusion coefficients multiplied by different dimensionless factors.  Two, in which the multiplicative factor is independent of $c_s$, are consistent with $c_s^{-4/5}$ scaling if the disk is isothermal.  In the first, labeled $R_{T1}$ the transition radius is proportional to $\alpha^{2/3}$ because, following \citet{PP83}, they supposed that the diffusion coefficient was multiplied by $1/\alpha$. The $R_{BP}$ formula, which assumes that radial mixing occurs at the same rate as accretion, leads to a diffusion coefficient multiplied by $\alpha$, so that the transition radius is $\propto \alpha^{-2/3}$.   In the third formula, $R_{T2}$, the diffusion is assumed to cease when the transition radius is small enough that $\alpha < H/r$; this has the effect of changing the sound speed scaling and removing any $\alpha$-dependence from the expression for the transition front. Thus, the factors that convert the sound speed proportionality into an equality are important.  When there is no shear viscosity at all (as in both our MHD and HD simulations), no role for $\alpha$  remains: the stresses relevant to alignment are neither viscous nor directly related to the accretion stress. In our formulation, $\alpha$ never enters the calculation; perhaps unsurprisingly, the multiplicative factor is therefore a constant $\sim O(1)$ for both MHD and HD models.

 \citet{O99}, using quasilinear methods, developed an effective diffusion coefficient theory for warp propagation. His expression for the dimensionless factor multiplying $c_s^2/\Omega$, when evaluated for the approximate conditions of our MHD simulations, is also $\simeq 2$.  Although this agreement is interesting, the reason for it is unclear, as this theory depends in an essential way on the assumption of an isotropic $\alpha$ viscosity, and the value of the diffusion constant in turn depends on $\alpha$.  We previously demonstrated that the assumption of an isotropic $\alpha$ is not supported by actual MHD calculations \citep{SKH13b}.  Moreover, our results found very similar behavior for both MHD disks and for thin disks in pure HD, i.e., with zero accretion stress so that $\alpha = 0$.   This indicates that value of the accretion stress does not significantly matter to the effective diffusion coefficient predicting the time-steady alignment front location.  In addition, \cite{O99} is a time-dependent theory and, as we have already discussed, our simulations do not support the use of a diffusion model for time-dependent properties.

To close, we point out another reason that may explain the limitations of diffusion in this context.  The alignment front's oscillation in position is closely associated with the state of the precession phase profile.   Whereas diffusion models are intrinsically local, the precession phase profile arises from global effects having to do with the radial propagation of bending waves \citep[e.g.,][]{PapTerq95,Larwood96}.   This global influence is another reason why diffusion models may not be adequate for describing alignment dynamics.   In the next subsection we will expand upon the importance of the precession phase profile.

\subsection{Precession phase gradients}

Lense-Thirring torques (and qualitatively similar torques like those produced by Newtonian quadrupoles) are purely precessional: the direction of the torque is precisely perpendicular to the angular momentum direction of the matter upon which the torque acts; the torque is also perpendicular to the angular momentum of the central object (black hole spin for Lense-Thirring torques, the orbital axis when a Newtonian quadrupole is due to a binary).  For this reason, if all the matter has the same angular momentum orientation, i.e., it forms a flat disk in which all orbits share the same orbital axis for all times, these torques can {\it never} cause alignment.  Instead, they force the matter's angular momentum to precess around the direction of the instigating angular momentum without any change in angular momentum magnitude.  The only way external torque can lead to alignment is if the torque that is identically perpendicular to the local angular momentum where it is delivered is then transferred to a place where it is no longer perpendicular; this can occur only if the precession phase varies with radius within the disk.  Moreover, this transfer causes alignment rather than enhanced misalignment only if the precession phase of the material where the angular momentum is ultimately deposited lags behind the precession phase of the material suffering the initial torque.

In these simulations, both the MHD and the HD, the significance of precession phase gradients has been made still clearer.  Alignment front progress is often non-monotonic, with both overshoots and retreats before the front becomes stationary.  The turn from outward motion to inward is, without exception, associated with a diminution in the precession phase gradient just outside the instantaneous position of the front; reversal of inward motion to outward is equally strongly associated with restoration of a negative radial precession phase gradient.   Thus, the regulation of precession phase gradients is a central element of alignment dynamics, and these gradients are controlled by two mechanisms triggered by warps: radial flows and bending waves.

Conversely, alignment fronts become stationary when the gradient in alignment angle is sufficient for inward and outward mixing to balance, while the precession of the disk well outside the alignment front is very nearly solid-body.  The precession rate for the outer disk under these circumstances is close to the Lense-Thirring rate at the radius of the mean angular momentum, i.e., $r_j = \int_{r_T} \, dr \, r^2 \ell(r)\Sigma(r)/\int_{r_T} \, dr \, r \ell(r) \Sigma(r)$.  Here $\ell$ is the local specific angular momentum and the lower limit of the integral is $r_T$ so as to include only the disk outside the alignment front.  The identification is quantitatively quite close: in Thin-H, for example, the outer disk precession rate matches the Lense-Thirring rate at $r \approx 17$, while $r_j$ defined by the integral is $\approx 19$.  The contrast between Thin-H and Big-H is particularly striking in this context; the existence of the extended outer disk in Big-H has a profound effect on the resulting solid-body precession rate.

This result is consistent with previous simulations, e.g., \cite{Fragile07},  who found that the final solid body precession rate was comparable to the value expected for the averaged angular momentum of their finite disk. We observe the same outcome for solid-body precession rates with both our ``small'' and ``large'' disks.  That the precession rate drops toward zero in our largest disk suggests a potentially interesting implication for disks in Nature that extend far beyond their alignment transition radius.

\subsection{Warp categorization}

\citet{PP83} argued, having assumed that all internal stresses in accretion disks are due to an hypothesized isotropic $\alpha$ viscosity, that warped disk behavior would exhibit two different regimes, depending on whether $\alpha$, the ratio of vertically-integrated accretion stress to vertically-integrated pressure, were greater or less than the disk aspect ratio $h/r$.  The regime in which $\alpha > h/r$ is the ``diffusive" regime, while its opposite is the ``bending wave" regime.  This characterization has become a widely adopted working hypothesis for disk alignment.

Whether this characterization has merit rests in part on the question ``What is $\alpha$?''   \cite{SS73} introduced $\alpha$ to parameterize the $r$--$\phi$ stress that is responsible for transporting angular momentum and driving accretion,  choosing to measure it in units of the pressure $P$.  But as \cite{Pringle92} observed, the stress component responsible for the damping of a warp and subsequent alignment is distinct from that responsible for accretion and may not have the same value of, or dependence on, $\alpha$.   Now that it is well-established that internal disk stresses driving accretion are primarily Maxwell stresses arising from MRI-driven MHD turbulence \citep{bh98},  the issue need no longer be a matter of speculation and parameterization.  Studies of MRI-induced turbulence have shown that MHD stresses acting on the shear flows induced by disk warps (the $r$--$z$ component of the stress tensor) bear no resemblance to an isotropic viscosity \citep{SKH13b,Fragile14a}.  This  conclusion that warp evolution has nothing to do with an ``isotropic $\alpha$ viscosity" is reinforced by the results from our HD simulations, which align despite being wholly inviscid ($\alpha = 0$).  
The dissipation/transport mechanism governing alignment is instead shocks and pressure-driven flows.  Given these observations, one might wonder what physical meaning remains to a distinction between diffusive and wavelike regimes determined by the value of $\alpha$.   The simulations reported here deepen the thrust of that question.

 As shown by \citet{BP99}, it is possible to reinterpret $\alpha$ as a measure of accretion stress caused by MHD turbulence, but only in reference to long-term time-averages of vertically-integrated quantities.  The $\alpha$ parameter is then the ratio of the integrated and averaged stress to the similarly integrated and averaged pressure.  As \citet{BP99} also showed, this identification breaks down badly when taken locally with respect to either time or spatial position.  In that spirit, we can measure the accretion stress in our simulations
 (the $r$-$\phi$ component of the Maxwell stress tensor; see eq.~\ref{eq:alpha})\footnote{The total stress includes an additional part from the Reynolds stress due to the MHD turbulence.  Typically the Maxwell stress is larger than the Reynolds stress by a factor of 3--4 \citep{bh98} and inclusion of the Reynolds stress would increase the effective value of $\alpha$ accordingly.  This has no qualitative impact on our conclusions here.} in units of pressure, even though its underlying mechanism is not at all viscous.  In our earliest and hottest ($h/r \simeq 0.2$) simulation \citep{SKH13b}, the magnitude of the stress normalized in this fashion was $\approx 0.02$--0.04, considerably less than $h/r$.  In KH2015, whose aspect ratio at the fiducial radius was 0.1, the $\alpha$ parameter was $\approx 0.03$--0.1, just a little bit smaller than  $h/r$.  In our new simulation with a fiducial $h/r = 0.05$ (High-thin), the stress parameter was $\simeq 0.2$, so that $\alpha$ was several times larger than $h/r$.  Finally, in our thinnest simulation, in which the fiducial $h/r = 0.035$ (V-thin), $\alpha$ was $\simeq 0.1$--0.15, also several times larger than $h/r$.  In other words, when the disks were relatively thick, $h/r$ was significantly larger than the normalized accretion stress, while in the thinner disks $h/r$ was rather smaller than the normalized accretion stress.  Strikingly, despite the fact that they span what had been predicted to be a marker of qualitative change, we have seen very little contrast in the way these disks align beyond that which is attributable to the sound speed (and $h/r$) itself.  Further, in the thinnest disks the alignment in the MHD disk is qualitatively similar to that seen in the HD disk, despite the fact that the HD disks have $\alpha = 0$, and would never be in the diffusive regime, at least as normally defined.  Put another way, the coefficient term accompanying the $c_s^{-4/5}$ dependence in the diffusion model at most depends very weakly on internal accretion stresses ($\alpha$), at least within the ranges of values explored here. It appears that the posited regime distinction based on $\alpha$ does not, in fact, matter very much.

On the other hand, we have seen that bending waves in purely hydrodynamic disks change from propagating waves to
strongly-damped waves as a function of a {\it different} dimensionless parameter, $\hat\psi \equiv |\partial \hat\ell /\partial \ln r|/(h/r)$.  When the warp is such that this ratio is relatively small, as in KH2015-H or the HD simulation of \cite{SKH13a}, bending waves propagate without dissipation, and can therefore travel long distances through the disk while losing little energy or angular momentum.  However, when $\hat\psi$ is large (as in Thin-H and V-thin-H), disk bends cause strong local pressure-driven flows and shocks, which quickly drain energy from any associated bending waves \citep{SKH13a}.    \cite{NP99}, in their study of bending waves, also observed that the amplitude of the warp relative to $h/r$ determined whether waves could propagate or were damped.  Unlike a condition based on  $\alpha$, this $\hat\psi$ condition is based on actual physical quantities.  Whether $\hat\psi$ is large or small determines the ability of the waves to bring the disk into a state of solid-body precession, a consequence of which is to bring to an end any further alignment.  To reiterate, the quantity determining that critical value is the warp rate, as opposed to a viscosity coefficient, and the mechanism is different, namely the disappearance of shocks limiting the propagation of bending waves, as opposed to the bending wave radial crossing time becoming shorter than the nominal ``viscous" damping rate of the waves.
 
The contrast between KH2015 (alignment) and KH2015-H (no alignment) and the MHD-HD pair in \cite{SKH13b} show that even when $\hat\psi$ is small,  and non-diffusive linear wave propagation would be expected, the presence of MHD turbulence can nevertheless disrupt the wave propagation through the disk, helping to preserve the precession phase contrast that allows for alignment.  Even in the ``Thin'' and ``Very Thin'' paired models where the HD model did align, the MHD models had larger transition front radii and extended alignment.  Thus, the accretion stress due to the MHD turbulence does affect alignment, but its role is indirect and accomplished by helping to delay the onset of solid-body precession and preserve differential precession rather than through direct diffusion of the disk warp. 
 
\subsection{Is MHD necessary?}

We have shown that in moderately thick disks (those of \cite{SKH13b} and KH2015, $h/r \gtrsim 0.1$), the alignment behavior in MHD is strikingly different than in HD.   On the other hand, the contrast in thinner disks ($h/r \lesssim 0.05$), appears to be much weaker, although still present.  The alignment fronts in the MHD simulations High-thin and V-thin continue outward beyond the time when their HD counterparts stop and retreat.   The reduced-resolution MHD model Low-thin has an intermediate behavior between models High-thin and Thin-H.  Finally, it is worth noting that the values of the quality metrics in the MHD simulations suggest that all the MHD models were somewhat under-resolved and that with better resolution the contrast between MHD and HD for thin disks may increase.

These observations lead to two suggestions.  The first is the one just made at the end of the previous subsection, that progression toward solid-body precession is hampered when the disk temperature is cool enough to make the bending waves induced by differential precession nonlinear.  Consequently, cool HD disks are able to align, even though they remain laminar.  Note that the criterion for being ``cool enough" is related to the initially-imposed warp rate $\psi$: more strongly warped disks can produce nonlinear bending waves for larger values of $h/r$, and therefore align even when they are not subject to MHD turbulence.   The second is that, because such thin disks are likely to be the norm rather than the exception in radiatively-efficient accretion flows, alignment may take place in disks immune to MHD effects (if, for example, they are too weakly-ionized for ideal MHD to be a good approximation) as well as in disks where MHD turbulence is strong.  In other words, disks possess two solutions to the problem of preserving their precession phase gradients: MRI-driven turbulence can disrupt bending wave passage whether the bending waves are linear or nonlinear; alternatively, in the absence of turbulence, if the bending waves are launched in a nonlinear state, shocks can limit their propagation.  The two solutions differ in the path through which disks reach stationary alignment states, but the character of those stationary states can be similar.

In this context, it is also worth noting that the $\alpha$-viscosity SPH simulation of \citet{Nealon2016} evolved very similarly to our MHD simulation KH2015.
Because we have already shown \citep{SKH13b} that MHD stresses do extremely little to restrain $r$-$z$ shear, a role that is at the heart of the traditional \cite{PP83} picture,  the agreement between these two simulations suggests that the detailed differences between MHD stresses and those resulting from the specific combination of explicit viscosity and numerical diffusivity present in the SPH simulation must not matter significantly in this particular instance. \footnote{The SPH algorithm requires an artificial viscosity.  If it is to take the form of an ``isotropic $\alpha$ viscosity", it entails a sizable bulk viscosity and a fixed ratio between the SPH smoothing length and the disk's vertical scale height \citep{Lodato10}.  This last constraint can lead to inadequate averaging in the smoothing volumes in low density regions, such as those away from the disk midplane---where much of the radial motion characteristic of warped disk evolution takes place.}

\section{Conclusions}

We have carried out a series of time-dependent, three-dimensional simulations in both HD and MHD designed to investigate the effect of sound speed on the alignment process in disks subject to Lense-Thirring torque.  From these results we have constructed a simple model for the position of the steady-state alignment front in which outward transport of angular momentum delivered by the Lense-Thirring torque is balanced by diffusive inward transport controlled by radial pressure gradients, gravity, and shocks.   This model leads to a sound speed scaling prescription for the location of the steady-state alignment transition, $r_T \propto c_s^{-4/5}$.  In doing so we have achieved, at least tentatively, a major goal of this subject: the ability to predict the location of the stationary alignment front in disks subjected to Lense-Thirring torques.  Compared to the traditional picture, in which radial mixing flows are regulated by an hypothesized ``isotropic $\alpha$ viscosity", the steady-state alignment fronts we predict---and find in explicit MHD simulations---are at radii substantially larger.  Moreover, the numerical tests we presented here have confirmed that the magnitude of the effective warp diffusion coefficient is given by a constant order-unity number, independent of the level of accretion stresses. Our data, limited unfortunately to a small number of cases, suggests that the diffusion coefficient is $\simeq 2 c_s^2/\Omega$.

Our ability to span at least some dynamic range in sound speed (about a factor of 3), and hence a range of disk thickness $h/r$,  has also permitted us to test a rubric that has long served as a fundamental guide to the field: the distinction between ``diffusive" and ``bending wave" regimes of disk alignment.  The traditional discriminant between the two is the amplitude of the viscous parameter $\alpha$ relative to $h/r$.   Based as it is on an extension of the $\alpha$ model, a model itself founded on dimensional analysis more than specific physical mechanisms, it may perhaps not be too surprising that this rubric fails when confronted with a direct test.   

Although we have laid out a ``diffusion model" for the location of the stationary alignment front, our orientation diffusion model differs in several significant ways from the traditional picture.   First, in our model, the internal accretion stress level is set by disk dynamics, and is not an adjustable free parameter; whether the disk aspect ratio is large or small compared to the normalized amplitude of the accretion stress seems to have little impact on alignment behavior.   Particularly notable is the similarity in alignment in sufficiently thin disks between turbulent MHD and HD disks, even though the latter are essentially inviscid, at least with respect to any internal accretion stress.   While the accretion stress does have an indirect, and possibly significant, role to play (see below), it is not primarily responsible for alignment, or for establishing the alignment radius $r_T$. 

Second, our diffusion model applies {\it only} to the prediction of steady-state alignment properties; indeed, it fails to predict significant elements of time-dependent behavior, such as overshoot and oscillation in the location of the alignment front.   In fact, it is possible that {\it no} diffusion model can adequately describe time-dependent alignment behavior.  Work over a number of years \citep{NP99,Lodato10,SKH13a} has raised suspicions that this might be the case.  Here we emphasize another reason why this may be so: changing alignment by making use of external precessional torques requires the correct precession phase gradient. This is an intrinsically global property because it involves communication over order-unity radial contrasts, whereas diffusion models refer only to local gradients in orientation.  Traditional ``diffusive regime'' solutions automatically impose strict locality by asserting that bending waves are damped after having traveled a distance that is a small fraction of their launch-point's radius.

Third, the ability of bending waves to propagate is not limited by a phenomenological viscosity, but by either of two physical mechanisms.   In MHD, their propagation is hampered by the turbulence created by the MRI.   In HD, the amplitude of the warp $\psi$ relative to $h/r$ determines the character of their propagation.  At low amplitudes, HD warps are able to propagate radially over extensive distances without losses.  At higher, nonlinear amplitudes, the waves are stymied by pressure gradient-driven Reynolds stresses and shocks.  In either MHD or HD, the ability of these waves to travel within the disk determines whether and when solid-body precession is established.

Lastly, we note that although MRI-driven turbulence should exist in all of the many sorts of disks in which the conductivity is high enough to support ideal MHD behavior, the strong damping of nonlinear bending waves provides Nature with a second way to prevent bending waves from spoiling the precession phase gradients necessary for alignment.  All that is necessary in laminar hydrodynamic disks is for the initial warp rate $|\partial \hat\ell/\partial \ln r|$ be large enough that it significantly exceeds $h/r$.   In the complete absence of internal accretion stress, there is a distinct regime in which a diffusion-like process determines the location of the steady-state alignment front, but its boundary is determined by the disk aspect ratio relative to the warp amplitude, not the accretion stress.

 \section*{Acknowledgements}

This work was partially supported under National Science Foundation grant
AST-0908869 (JFH), NASA grant NNX14AB43G (JHK and JFH), and NSF grants AST-1516299 and AST-1715032 (JHK).  This work was also partially supported by a grant from the Simons Foundation (559794, JHU).
The authors acknowledge the Texas Advanced Computing Center (TACC) at The University of Texas at Austin for providing high performance computing resources that have contributed to the research results reported within this paper.
The National Science Foundation supported this research through XSEDE allocation TG-MCA95C003.  This research is also part of the Blue Waters sustained-petascale computing project, which is supported by the National Science Foundation (awards OCI-0725070 and ACI-1238993) and the state of Illinois. Blue Waters is a joint effort of the University of Illinois at Urbana-Champaign and its National Center for Supercomputing Applications.  The Blue Waters allocation was obtained through an award from the Great Lakes Consortium for Petascale Computation.  We thank Scott Noble for assistance with Blue Waters simulations. Additional simulations were carried out on the {\it Rivanna} high performance cluster at the University of Virginia with support from the Advanced Research Computing Services group.

\bibliography{Bib}

\end{document}